\documentclass[aps,prd,onecolumn,preprintnumbers,floatfix,showpacs,
nofootinbib,superscriptaddress,notitlepage]{revtex4-1}

\usepackage[justification=raggedright,singlelinecheck=false]{caption}
\usepackage{amsmath}           %
\usepackage{amssymb}           %
\usepackage{color}             %
\usepackage{bbm}               %
\usepackage{braket}            %
\usepackage{xspace}            %
\usepackage{mciteplus}         %
\usepackage{mathtools}         %
\usepackage{graphicx}          %
\usepackage{slashed}           %
\usepackage{enumerate}         %
\usepackage[export]{adjustbox} %
\usepackage[english]{babel}    %
\usepackage[utf8]{inputenc}    %
\usepackage{hyperref}          %
\usepackage{booktabs}
\usepackage{tabularx}
\usepackage{float}
\usepackage{subcaption}
\usepackage[normalem]{ulem}

\clearpage{}

\renewcommand{\k}{\mathbf{k}}
\newcommand{\p}{\mathbf{p}}
\newcommand{\q}{\mathbf{q}}

\renewcommand{\P}{\mathbf{P}}

\newcommand{\1}{\mathbf{1}}
\newcommand{\2}{\mathbf{2}}

\newcommand{\Ac}{\mathcal{A}}
\newcommand{\Actwo}{\Ac}

\newcommand{\Gc}{\mathcal{G}}

\newcommand{\Jc}{\mathcal{J}}

\newcommand{\Lc}{\mathcal{L}}
\newcommand{\Mc}{\mathcal{M}}

\newcommand{\Oc}{\mathcal{O}}

\newcommand{\Rc}{\mathcal{R}}

\newcommand{\Tc}{\mathcal{T}}

\newcommand{\Wc}{\mathcal{W}}

\newcommand{\NR}{{\rm{NR}}}

\newcommand{\abs}[1]{\ensuremath{\lvert#1\rvert}}
\newcommand{\ev}[1]{\ensuremath{\langle#1\rangle}}

\newcommand{\df}{\textrm{df}}
\newcommand{\nn}{\nonumber}
\newcommand{\diff}{\textrm{d}}

\newcommand{\beq}{\begin{equation}}
\newcommand{\eeq}{\end{equation}}

\newcommand{\kev}{\ensuremath{{\mathrm{\,ke\kern -0.1em V}}}\xspace}
\newcommand{\mev}{\ensuremath{{\mathrm{\,Me\kern -0.1em V}}}\xspace}
\newcommand{\gev}{\ensuremath{{\mathrm{\,Ge\kern -0.1em V}}}\xspace}
\newcommand{\tev}{\ensuremath{{\mathrm{\,Te\kern -0.1em V}}}\xspace}

\usepackage{orcidlink}

\usepackage{mfirstuc} %
\newcommand{\addReviewer}[2]{
  \expandafter\newcommand\csname #1\endcsname[1]{{\bf \color{#2} \capitalisewords{#1}:\,##1}}
  \expandafter\newcommand\csname #1cor\endcsname[2]{{\color{#2} \capitalisewords{#1}:\,\st{##1}{\bf ##2}}}
  \expandafter\newcommand\csname #1color\endcsname{#2}
}
\usepackage{soul,color}
\definecolor{chromeyellow}{rgb}{1.0, 0.65, 0.0}
\definecolor{DodgeBlue}{rgb}{0.118, 0.565,1.000}
\definecolor{asparagus}{rgb}{0.53, 0.66, 0.42}
\definecolor{cadmiumgreen}{rgb}{0.0, 0.42, 0.24}
\addReviewer{andrew}{cadmiumgreen}

\clearpage{}

\hypersetup{ 
    pdfnewwindow=true,      %
    colorlinks=true,       %
    linkcolor=blue,          %
    citecolor=blue,        %
    filecolor=blue,      %
    urlcolor=blue        %
} 
\definecolor{jlab_red}{RGB}{192,39,45}
\definecolor{jlab_orange}{RGB}{249,102,0}
\definecolor{jlab_blue}{RGB}{47,122,121}
\definecolor{jlab_green}{RGB}{65,125,10}
\definecolor{carolinablue}{cmyk}{0.60,0.19,0.01,0.01}

\newcommand{\ck}[2]{
  \if\relax\detokenize{#1}\relax\else
    \bgroup
    \markoverwith{\textcolor{carolinablue}{\rule[0.5ex]{2pt}{0.4pt}}}%
    \ULon{#1}%
    \egroup
  \fi
  \if\relax\detokenize{#1}\relax\else
    \if\relax\detokenize{#2}\relax\else
      \space
    \fi
  \fi
  \if\relax\detokenize{#2}\relax\else
    {\sf \color{carolinablue} #2}
  \fi
}

\newcommand{\unc}{Department of Physics \& Astronomy,
University of North Carolina, 
Chapel Hill, NC, 27599, USA}

\newcommand{\wm}{Department of Physics,
William \& Mary,
Williamsburg, VA 23187, USA}

\newcommand{\ucb}{Department of Physics,
University of California,
Berkeley, CA 94720, USA}

\newcommand{\lbnl}{Nuclear Science Division,
Lawrence Berkeley National Laboratory, Berkeley,
CA 94720, USA}

\newcommand{\umd}{Department of Physics,
University of Maryland, 
College Park, MD, 20742, USA}
\begin{document}
\title{
Resolving the structure of bound states 
\\
using lattice quantum field theories
}
%

%	AUTHORS
%%%%%%%%%%%%%%%%%%%%%%%%%%%%%%%%%%%%

%%%%%%%%%%
\author{Joseph Moscoso~\orcidlink{0000-0002-7227-9165}}
 \email[e-mail: ]{jmoscoso@umd.edu}
\affiliation{\umd}
\affiliation{\unc}
\affiliation{\ucb}
\affiliation{\lbnl}
%%%%%%%%%%

%%%%%%%%%%
\author{Felipe G. Ortega-Gama~\orcidlink{0000-0001-8453-9481}}
\email[e-mail: ]{fgortegagama@berkeley.edu}
\affiliation{\ucb}
\affiliation{\lbnl}
%%%%%%%%%%

%%%%%%%%%%
\author{Ra\'ul A. Brice\~no~\orcidlink{0000-0003-1109-1473}}
\email[e-mail: ]{rbriceno@berkeley.edu}
\affiliation{\ucb}
\affiliation{\lbnl}
%%%%%%%%%%

%%%%%%%%%%
\author{Andrew~W.~Jackura~\orcidlink{0000-0002-3249-5410}}
\email[e-mail: ]{awjackura@wm.edu}
\affiliation{\wm}
%%%%%%%%%%

%%%%%%%%%%
\author{Charles Kacir~\orcidlink{0009-0008-2736-4599}}
\email[e-mail: ]{ckacir@unc.edu}
\affiliation{\unc}
%%%%%%%%%%

%%%%%%%%%%
\author{Amy N. Nicholson}
\email[e-mail: ]{annichol@email.unc.edu}
\affiliation{\unc}
\affiliation{\lbnl}
%%%%%%%%%%

%
\begin{abstract}
This work presents the first lattice calculation of a two-to-two particle matrix element of a local current. This exploratory calculation is performed using a leading-order pionless effective field theory of two nucleons in a finite 3D spatial volume, where the Hamiltonian can be diagonalized exactly for moderate volumes. By considering a range of couplings where the theory supports a deuteron-like bound state, we determine the finite-volume spectra and matrix elements of the conserved local vector current. Using the L\"uscher formalism, we constrain the infinite-volume, purely hadronic amplitude for this theory. Using previously derived formalism, we then map the finite-volume matrix elements to scattering amplitudes describing a reaction coupling two-particle states via a current insertion, $\2+\Jc \to \2$. We then use a recently derived relation between this class of amplitudes and the bound-state elastic form factor to directly constrain the infinite-volume form factor. By varying over a range of values of the coupling of the theory, we explore the effects of this analysis for deep-bound states and shallow-bound states. We reproduce the expected result that for deep bound states, the finite-volume formalism is largely unnecessary, while for shallow bound states, it is absolutely critical to obtain a sensible result. We present a detailed outline of the analysis of this class of matrix elements, including the determination of the charge radius of the bound state. In the shallow bound state limit, we find good agreement with the prediction stemming from the anomalous threshold. 
\end{abstract}
\date{\today}
\maketitle

\section{Introduction}\label{sec:intro}
A long-standing goal of nuclear physics is to establish reliable predictions of the properties and reactions of nuclei directly from Quantum Chromodynamics (QCD). Constraining the dynamics of quarks and gluons, and how they bind together, first into protons and neutrons, and then into nuclei, is a notable theoretical challenge due to the non-perturbative nature of QCD. However, precise theoretical input is needed, for example, to constrain the dynamics of nuclear interactions, which itself is critical for having a quantitative understanding of the most basic nuclear reactions driving the universe's evolution~\cite{Iocco_2009,Adelberger_2011,RevModPhys.88.015004}. Additionally, characterizing the hadronic responses to electroweak probes is necessary for understanding a large number of experimental searches constraining beyond-Standard-Model (BSM) physics scenarios, like long-baseline neutrino experiments, which use the nucleus as a laboratory to probe the nature of neutrinos~\cite{Alvarez_Ruso_2018,Kronfeld_2019,Arguelles:2019xgp,DUNE:2020fgq,DUNE:2022aul}. Determining the relevant electroweak matrix elements of these processes is necessary to enhance the discovery potential of such experiments and to provide insight into their nature.

First-principles calculations of the structure and properties of nucleons and bound nuclei are possible with lattice QCD (LQCD), a numerical approach that yields quantitative predictions for hadronic observables. In the last decade, we have witnessed significant progress in the determination of elastic electroweak form factors of the nucleon~\cite{FlavourLatticeAveragingGroupFLAG:2024oxs,Gupta:2024qip,Tsuji:2023llh,Aoki:2025taf,Alexandrou:2020okk,Alexandrou:2025vto,Barone:2025rye,Hackett:2023rif, Hall:2025ytt}, as well as parton distribution functions~\cite{HadStruc:2021qdf,HadStruc:2024rix,Lin:2021brq,Alexandrou:2019ali,Bhattacharya:2023ays,Bhattacharya:2024wtg,Chu:2025kew}, which are starting to paint a detailed picture of its internal structure. However, understanding the internal dynamics of light nuclei with LQCD is still in its infancy due to the complexity of the calculations and the amount of computational resources required for the extraction of baryon correlation functions, which are plagued by a severe signal-to-noise problem~\cite{Lepage:1989hd,Wagman_2017}. This signal-to-noise problem has resulted in the fact that, although we have known formally how to constrain $NN$ scattering amplitudes from LQCD for over a decade~\cite{Ishii:2006ec,Aoki:2009ji,Luscher:1990ux,Briceno:2013lba}, it has not been until recently that several groups using complementary techniques have obtained statistically consistent $NN$ scattering amplitudes at exploratory heavy quark masses~\cite{bulava2025dinucleonsformboundstates,H_rz_2021,PhysRevD.107.094508,Detmold:2024iwz,Inoue_2012,Francis_2019}, corresponding to $m_\pi\sim 700-800$~MeV.
All of these studies point to the fact that at these heavy quark masses, the isoscalar and isovector $NN$ systems have a shallow virtual bound state. This is in contrast with the experimental situation, where it is well-known that the $nn$ channel has a virtual bound state, namely the dineutron, while the isoscalar channel contains the deuteron, which appears as a real bound state. Understanding the evolution of these states as the pion mass is reduced to the physical point is of paramount importance for controlling LQCD calculations of light nuclei.

Now that we have reliable methods for computing the purely hadronic $NN$ amplitudes, it is a reasonable time to investigate the feasibility of accessing the structural information of these states. There are presently two classes of formalisms for performing these calculations in the literature. The first uses the fact that at low energies, one can construct an effective field theory (EFT) to describe both finite- and infinite-volume observables in terms of volume-independent low-energy coefficients~\cite{Detmold:2004qn, Briceno:2012yi,Lozano:2022kfz,Davoudi:2020xdv}. This allows for an indirect matching of, for example, finite-volume matrix elements and infinite-volume electroweak processes, which holds up to a finite order in the expansion of the EFT. Consequently, the resulting amplitudes can only be as precise as the EFT and the order chosen in performing the analysis. Alternatively, one can derive a more universal class of correspondences between finite-volume matrix elements of local currents and infinite-volume amplitudes. These correspondences can be derived using all-orders perturbation theory~\cite{Briceno:2015tza, Baroni:2018iau, Briceno:2020vgp}. Furthermore, these correspondences can be obtained for large classes of reactions using a minimal set of kinematic assumptions. As a result, one can imagine implementing these for studying, for example,
\begin{itemize}
    \item coherent neutrino deuteron scattering, $\nu_l d \rightarrow \nu_l d$,
    \item photodisintegration of the deuteron, $\gamma^* d \rightarrow np$,
    \item elastic form factors of the deuteron, $\gamma d \rightarrow \gamma d$.
\end{itemize}

Currently, there are two key assumptions made in the derivation of Refs.~\cite{Briceno:2015tza, Baroni:2018iau, Briceno:2020vgp} that limit their applicability. First, the formalism only holds for kinematics, where only two-particle systems can go on shell. Second, it assumes that all particles involved are spinless. Therefore, until this formalism is generalized to incorporate the intrinsic spin of nucleons, it can only be rigorously used for kinematics where spin effects are negligible.

\begin{figure}[t]
    \centering
    \includegraphics[width=0.85\linewidth]{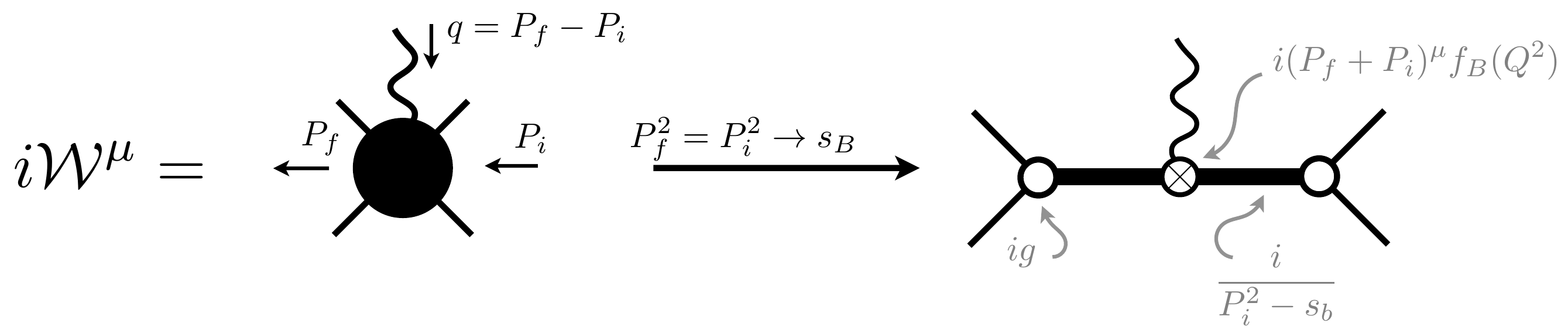}
     \caption{Shown is a qualitative depiction of the $\2+\Jc^\mu\to\2$ amplitude, $\Wc^\mu$, and its relation to the bound state form factor, $f_B$. Note $Q^2=-(P_f-P_i)^2$.
    }
    \label{fig:bs_ff}
\end{figure}

In brief, the formalism presented in Refs.~\cite{Briceno:2015tza, Baroni:2018iau, Briceno:2020vgp} first relates the finite-volume two-particle matrix elements $\langle\2 | \Jc|\2\rangle$, where $\Jc$ is a local current, to an infinite-volume amplitude $\2+\Jc\to \2$, denoted as $\mathcal{W}$. These classes of amplitudes are largely unexplored in the literature, but they contain a tremendous amount of information. For example, if these processes support bound states or resonances, one can obtain elastic factors and inelastic transition amplitudes from the pole singularities of these amplitudes. This procedure, which is qualitatively depicted in Fig.~\ref{fig:bs_ff}, is the main focus of this work. 

This formalism is absolutely critical if one wishes to access $\2+\Jc\to \2$ amplitudes via LQCD. This is because scattering/reactions are strictly inaccessible in finite volumes, where LQCD calculations are necessarily performed. This formalism directly maps finite-volume observables to infinite-volume observables. This mapping is exact up to exponentially suppressed errors that scale as $e^{-mL}$, where $L$ is the spatial extent of the lattice and $m$ is the mass gap of the theory. This formalism builds on previous formalisms presented in Refs.~\cite{Lellouch:2000,Christ:2005gi, Brice_o_2015a, Brice_o_2015b, Agadjanov_2016} for studying a simpler class of reactions, which are of the type $ \Jc \to \2$ or $\1 +\Jc \to \2$. These other formalisms have been used in lattice QCD studies of an increasingly large number of electroweak transition amplitudes~\cite{Brice_o_2016, Alexandrou_2018,PhysRevD.106.114513,Chen20221HI,Feng:2014gba,Andersen:2018mau,Erben:2019nmx,Radhakrishnan:2022ubg, ortegagama2024timelikemesonformfactorselastic}. These studies provide promising progress toward the consideration of more complicated electroweak transitions, with the most natural case being of the form $\2+\Jc\to \2$.

Before this work, several non-trivial checks have been performed on the formalism presented in Refs.~\cite{Briceno:2015tza, Baroni:2018iau, Briceno:2020vgp}. These include studying the limit where a two-body system supports a bound state and when the amplitude satisfies the Ward-Takahashi identity. In Ref.~\cite{Briceno:2019nns}, it was shown that the correction provided by the formalism becomes exponentially small as the binding energy increases, agreeing with the expectation that matrix elements of stable states have negligible volume dependence. It was also shown that the amplitude extracted satisfies constraints placed by the Ward-Takahashi identity when a conserved vector current is considered.
In a parallel study, it was shown that the formalism reproduces the expectations from perturbation theory~\cite{Briceno:2020xxs}. In particular, the non-perturbative formalism was used to describe a weak system, where the finite-volume spectrum and matrix elements could be evaluated perturbatively, and perfect agreement was found when perturbatively expanding the result of the formalism. Furthermore, it was shown that the volume scaling of matrix elements for a scalar current was consistent with the expectation from the Feynman-Hellmann theorem. 

In this work, we present the first implementation of this formalism for a lattice quantum field theory calculation~\footnote{For an ongoing parallel Monte Carlo calculation of this same class of reactions in the $O(3)$ non-linear sigma model in $1+1D$, we point the reader to Ref.~\cite{nlsm:2026}.}. In particular, we evaluate the finite-volume spectrum and matrix elements for the so-called pionless nuclear EFT ($\text{EFT}_{\slashed \pi}$)~\cite{Kaplan_1998_1, Kaplan_1998_2}.~\footnote{Lattice EFTs have been used to study the properties of light- and medium-mass nuclear systems~\cite{Lee:2025req, Epelbaum:2011md,Lahde:2013uqa,Shen:2024qzi,Shen:2021kqr,Ren:2023ued,Lu:2018bat}. Typically, these studies are performed at large enough volumes where finite-volume errors can be safely ignored.}  By considering moderately sized $3D$ volumes, we can diagonalize the Hamiltonian exactly. We tune the two-body interaction of the theory to support a bound state, and vary the parameter to study the matrix elements within deeply bound states, where we expect the formalism to be largely unnecessary, as well as increasingly shallow bound states. Although we have spectra and matrix elements below and above the two-particle threshold, in this analysis, we focus our attention on the energy levels below threshold. We find that for shallow bound states, without applying the finite-volume formalism, one would obtain form factors for the bound states that are multi-valued, and consequently violate analyticity. In other words, unless the binding momentum of the two-particle bound state is large in comparison to the inverse of the spatial extent of the lattice, the formalism is necessary and crucial.

The remainder of this work is organized as follows. In Sec.~\ref{sec:review}, we review the formalism presented in Refs.~\cite{Briceno:2015tza, Baroni:2018iau, Briceno:2020vgp} for studying two-body matrix elements and the corresponding amplitudes. In Sec.~\ref{sec:DiscPEFT}, we present our implementation of the lattice $\text{EFT}_{\slashed \pi}$. In Sec.~\ref{sec:results}, we present our analysis of the spectrum, matrix elements, and provide our determination of the infinite-volume form factors. Finally, in Sec.~\ref{sec:conclusion}, we conclude and discuss the outlook for the formalism. Additionally, we provide details regarding the projection procedure for the lattice in App.~\ref{app:proj} and the asymptotic form of the charge radius in App.~\ref{app:radius}

\section{Review of scattering and finite-volume formalism}
\label{sec:review}
Here we review the fully relativistic scattering formalism presented in Refs.~\cite{Briceno:2015tza,Baroni:2018iau,Briceno:2020vgp} for studying reactions where two-particle states are coupled via an external vector current $\Jc^{\mu}$, i.e.\ $\2+\Jc^\mu\to\2$ amplitudes. 
The key takeaway from these formalisms for the context of this work is that the fully relativistic form factor for a two-particle bound state $f_B$ satisfies the relationship
\begin{equation}\label{eq:ffb}
(P_f+P_i)^\mu\,f_B(Q^2) = g^2\left(\Actwo^{\mu}(P_f,P_i)
+(f_1(Q^2) + f_2(Q^2)) [(P_f+P_i)^\mu \Gc(s_B,Q^2,s_B) -2\Gc^\mu(P_f,P_i) ]
\right),
\end{equation}
where $P_{i/f}$ is the four momentum of the initial/final state, $Q^2=-(P_f-P_i)^2$ is the current virtuality, and $s_B=P_{i/f}^2$ is the mass squared of the bound state.
The different terms in this expression encode dynamical information of the two-particle system in the following way.
First, $g$ is the coupling of the two-particle scattering state to the bound state, which can be extracted from the two-particle scattering amplitude.
Second, the function $\Actwo^{\mu}(P_f,P_i)$ contains the short-distance dynamics between the two-particle system and the current.
Finally, $f_{1/2}(Q^2)$ stands for the individual form factor of each of the particles within the two-particle system, assuming that the two particles are distinguishable; while $\Gc$ and $\Gc^\mu$ are kinematic functions originating from the triangle diagram to be discussed later.

Below, we explain how this form factor can be obtained from the residue of the $\2+\Jc^\mu\to\2$ amplitude at the bound state pole~\cite{Briceno:2020vgp}. This is qualitatively depicted in Fig.~\ref{fig:bs_ff}. Furthermore, we explain how such an amplitude can be obtained from the analysis of finite-volume matrix elements of the current~\cite{Briceno:2015tza,Baroni:2018iau}, as well as the finite-volume spectrum. Before discussing these, we will need to define matrix elements of single-particle states as well as purely hadronic $\2\to\2$ amplitudes, since these are necessary inputs into the analysis. In what follows, we will assume the particles are degenerate but can be distinguished by the charge coupling to the external current. One of the particles will have a non-zero coupling to the current, while the other will have a 0 coupling (e.g.\ $\pi^+$ and $\pi^0$ or $p$ and $n$ in the isospin limit). In Sec.~\ref{sec:NR_lim}, we will discuss how one can match to non-relativistic QFTs.

\subsection{Single particle form factors}\label{sec:ff}

Let $\ket{P}_{\rm rel.}$ denote a \emph{relativistic} infinite-volume single particle state with four-momentum $P^\mu=(E,\textbf{P})$ with mass $M=\sqrt{P^2}$ and the standard relativistic normalization, 
\begin{align}
\label{eq:rel_basis}
\bra{P_f } {{P_i}\rangle}_{\rm rel.} = 2 E_f \, (2\pi)^3 \delta^3(\textbf{P}_f -  \textbf{P}_i)\,.
\end{align} 

With this, we can define matrix elements of a vector current, $\Jc^{\mu}(x)$, in terms of two Lorentz scalars (the form factors) which depend on the current virtuality $Q^2$,
\begin{equation}
\label{eq:F_decomp}
\bra{P_f } \Jc^\mu (x=0) \ket{P_i }_{\rm rel.} = (P_f + P_i)^\mu f_{1}(Q^2) +(P_f - P_i)^\mu f_{2}(Q^2)  \,.
\end{equation}
Requiring that this current is conserved, $\partial_\mu \Jc^{\mu}(x) = 0$, puts an additional constraint on the matrix element, which in momentum space can be written as
\begin{align}
(P_f - P_i)_\mu\bra{P_f } \Jc^\mu (0) \ket{P_i }_{\rm rel.} &= (P_f - P_i)_\mu(P_f + P_i)^\mu f_{1}(Q^2) +(P_f - P_i)^2 f_{2}(Q^2)  \,,
\nn\\
&= -Q^2 f_{2}(Q^2) =0\,,
\end{align}
which must be true for any value of $Q^2$, implying that $f_{2}(Q^2)=0$. In summary, for a conserved current, we need a single form factor, and as a result, we drop the subscript,
\begin{equation}
\label{eq:1body_me}
\bra{P_f } \Jc^\mu (0) \ket{P_i }_{\rm rel.} = (P_f + P_i)^\mu f(Q^2) \,.
\end{equation}

The fundamental fields will be structureless in the non-relativistic EFT used in this work, so their charge form factor will be equal to 1 for all kinematic configurations. However, Eq.~\eqref{eq:1body_me} also holds for the bound states of two particles. These bound states dynamically acquire non-trivial structure and will, in general, have a $Q^2$ dependent form factor. In Sec.~\ref{sec:2Jto2}, we describe the two-hadron formalism of Ref.~\cite{Briceno:2020vgp} and its application to bound states.

\subsection{Relativistic \texorpdfstring{$\2\to \2$}{2→2} scattering}

Here, we are interested in a field theory that only has one open two-particle channel composed of non-identical degenerate particles of mass $M$ that only couple to the $\ell = 0$ partial wave. In this limit, the purely hadronic $\2\to \2$ scattering amplitude can be written in terms of a single function, namely the $\ell = 0$ scattering phase shift, which we label as $\delta$. This is a function of the center-of-momentum (CM) energy $ E^\star$, or equivalently, the Mandelstam variable $s=E^{\star 2}$. Without loss of generality, one can write the scattering amplitude $\Mc$ as
\begin{equation}\label{eq:scat_rel}
\Mc(s) =   \frac{1}{\rho(s) \cot\delta(s) - i\rho(s)},
\end{equation}
where 
\begin{align}
\label{eq:rhos}
\rho(s)=\frac{q^\star}{8\pi E^\star},
\end{align}
is the phase space factor for distinguishable particles, and $q^\star=\sqrt{s/4 -M^2}$ is the CM relative momentum of the two-particle system. 

A useful low-energy parametrization of the scattering amplitude that we utilize in this work is the effective range expansion (ERE), where $q^{\star} \cot\delta(s)$ is expanded near threshold as a polynomial in $q^{\star 2}$. In what follows, we will only consider terms that go to next-to-leading order (NLO) in this expansion,
\begin{equation}\label{eq:ERE}
q^{\star} \cot\delta^{\rm NLO}(s) = -\frac{1}{a} + \frac{1}{2} r  q^{\star 2}  \,,
\end{equation}
where $a$ and $r$ are respectively known as the scattering length and effective range. 

Furthermore, we will explore values of these parameters for which the scattering amplitude can acquire a bound state. This occurs when the denominator of Eq.~\eqref{eq:scat_rel} vanishes. Denoting the binding momentum as $\kappa$, this occurs for energies below threshold satisfying,
\begin{equation}\label{eq:BS_pole}
q^{\star} \, \cot\delta(s)  \big\rvert_{q^{\star} = i\kappa} + \kappa = 0 \,.
\end{equation}
Ultimately, we will need to know not just the pole location of the scattering amplitude, but also the residue at the pole. This is given by the behavior of $\Mc$ near the bound state pole, $s_B$
 \begin{align}\label{eq:M_BS_pole}
\Mc(s) &\sim \frac{(ig)^2}{s - s_B}  \,,
\nn\\
\Rightarrow
g&= \lim_{s\to s_B} \sqrt{(s_B-s)\Mc(s)}.
\end{align}
After a bit of algebra, one finds that $g$ can be written as~\cite{Briceno:2019nns},
\begin{equation}
\label{eq:g2}
    \frac{1}{g^2} = \frac{1}{64 \pi \,\kappa\, \sqrt{s_B}} \Big( 1 - 2 \kappa \frac{\partial}{\partial q^{\star2}} q^\star \cot \delta (s) \Big )_{s = s_B}. 
\end{equation}

\subsection{Relativistic \texorpdfstring{$\2+\Jc^\mu\to\2$}{2+J→2} amplitudes }\label{sec:2Jto2}
We now review the definition of the $\2+\Jc^\mu\to\2$ amplitudes~\cite{Briceno:2020vgp, Briceno:2019nns}. As for the purely hadronic process, we restrict our attention to the $\ell =0$ partial wave. Just like in the one-particle sector, discussed in Sec.~\ref{sec:ff}, these amplitudes have a matrix element definition 
\begin{equation}\label{eq:W_matrix}
\bra{P_f,p_f} \Jc^{\mu}(0) \ket{P_i, p_i}_{\textrm{conn.}} \equiv \Wc^{\mu}(P_f, P_i) \,,
\end{equation}
where $\Wc^{\mu}$ is the desired amplitude, $\ket{P_{i}, p_i}/\ket{P_f,p_f}$ are in/out two-particle states with total momenta $P_i/P_f$ where one of the two particles carries momentum $p_i/p_f$, respectively, and the subscript $\textrm{conn.}$ reminds us that only connected diagrams contribute to this amplitude.

Reference~\cite{Briceno:2020vgp} shows that this amplitude can be written in terms of two pieces. The first is a sum of terms with simple poles associated with propagators, $D$, of single particles coupling to the external current, and is completely constrained by sub-process quantities that may be determined independently of $\mathcal{W}^\mu$, i.e.\ the two particle scattering $\Mc$ and terms proportional to the single particle form factors, $w^{\mu}_{\mathrm{on}}\propto f_{1/2}(Q^2)$. The second term is known as $\Wc^{\mu}_{\df}$, where the subscript alludes to its \emph{divergence-free} nature, since it no longer has simple single-particle poles. As explained in Ref.~\cite{Briceno:2020vgp}, this still has softer kinematic singularities and possible dynamical pole singularities, associated with bound states or resonances. In short, $\Wc^{\mu}$ can be written as,
\begin{align}
\label{eq:Wmu}
i\Wc^{\mu}(P_f, P_i)
\equiv 
   \sum 
   \left\{
   iw^{\mu}_{\mathrm{on}}\,
   iD\,
   i\Mc
   \right\}
    +i\Wc^{\mu}_{\df} (P_f, P_i).
\end{align}
In the limit that two-particle states couple to $\ell=0$, the pole piece within the summation depends on the angles of the relative momentum, while $\Wc^{\mu}_{\df} $ does not.  The latter is the piece that is dynamically interesting since it can provide nontrivial information on the internal structure of the two-particle state, so it is the focus of our attention in this work.

We consider the specific case of two distinguishable degenerate particles, one neutral and the other charged, with no internal structure, i.e.\ with single-particle form factors $f_1(Q^2) =1$ and $f_2(Q^2)=0$. Projecting to $\ell =0$, the general expression found in Ref.~\cite{Briceno:2020vgp} for $\Wc^{\mu}_{\df}$ can be written as,
\begin{align}
\label{eq:Wdf_df}
\Wc^{\mu}_{\df} (P_f, P_i)
=
\Mc (P_f^2)
\left[\Actwo^\mu (P_f, P_i)
+
(P_f+P_i)^{\mu}\Gc(P_f^2,Q^2,P_i^2)
-2\Gc^{\mu}(P_f,P_i)
\right]\Mc (P_i^2),
\end{align}
where $\Gc$ and $\Gc^\mu$ are relativistic infinite-volume triangle functions,
\begin{align}
\label{eq:Gcal}
\Gc(P_f^2,Q^2,P_i^2)&=i\int \frac{d^4k}{(2\pi)^4}
D\big(k^2\big)\,D\big((P_f-k)^2\big)\,D\big((P_i-k)^2\big),\, 
\\
\label{eq:Gcalmu}
\Gc^{\mu}(P_f,P_i)&=i\int \frac{d^4k}{(2\pi)^4}
k^\mu \, D\big(k^2\big)\,D\big((P_f-k)^2\big)\,D\big((P_i-k)^2\big).\, 
\end{align}
in which $D(k^2)=(k^2-M^2+i\epsilon)^{-1}$ is the single-particle propagator.
These integrals are UV convergent and can be evaluated using standard methods. Reference~\cite{Baroni:2018iau} provides a general prescription for evaluating this class of integrals, and Appendix~A of Ref.~\cite{Briceno:2020vgp} gives a thorough discussion on the singularity structure of these functions.
A key point is that, in addition to square-root threshold singularities also present in $\Mc$, these functions exhibit anomalous threshold singularities that can be logarithmic or inverse square root depending on the kinematics.

The only function that has not already been defined in Eq.~\eqref{eq:Wdf_df} is $\Actwo^\mu$. This was identified in Ref.~\cite{Briceno:2020vgp} to be purely real and a non-singular function of the external kinematics but is otherwise unconstrained. 
For conserved vector currents like the one considered in this work, the $\Wc^\mu$ amplitude is subject to the Ward-Takahashi identity
\begin{align}
q_\mu \Wc^{\mu} = (P_f - P_i)_\mu\Wc^{\mu}  (P_f,P_i)=0.
\end{align}
This equation can be used to further restrict $\Actwo^\mu$, however, it is not sufficient to completely determine its functional form. 
The proper normalization of the bound state form factor, i.e.\ $f_B(0)=f_1(0)$, follows from the amplitude satisfying the Ward-Takahashi identity~\cite{Briceno:2019nns}.

Let us now consider the consequence of this formalism in the limit where a bound state is present at $P_i^2=P_f^2=s_B$. 
Beyond this, we can follow Refs.~\cite{Briceno:2020vgp, Briceno:2019nns} to define the bound state form factor, $f_B$, from $\Wc$. 

If there is a bound state pole in $\Mc$, then $\Wc_\df$ will have a pole associated with the initial and final state. Using the Lehmann-Symanzik-Zimmermann reduction procedure to get the form factor from the residue of $\Wc^\mu$, 
\begin{equation}
(P_f + P_i)^\mu f_{B}(Q^2 ) = 
    \lim_{P_{f/i}^2 \rightarrow s_B} \frac{P_f^2 - s_B}{g }  \Wc_{\text{df}}^\mu(P_f, P_i) \frac{P_i^2 - s_B}{g }  .
\end{equation}
Note that whether one uses $\Wc_{\text{df}}$ or $\Wc$ on the right-hand side of the equality, one arrives at the same result because the single pole piece of Eq.~\eqref{eq:Wmu} does not contribute.

By inserting Eq.~\eqref{eq:Wdf_df} in this expression and canceling the poles appearing in $\Mc$, given by Eq.~\eqref{eq:M_BS_pole}, one arrives at an expression for $f_B$ in terms of the previously introduced functions, 
\begin{align}
\label{eq:ffb_v0}
(P_f+P_i)^\mu f_B(Q^2)=g^2\left(\Actwo^\mu (P_f,P_i)
+(P_f+P_i)^\mu\Gc(s_B,Q^2, s_B)-{2 \Gc^{\mu}(P_f,P_i)}
\right),
\end{align}
where all functions are evaluated at the bound state energy, $s_B$. This finalizes the derivation of Eq.~\eqref{eq:ffb} quoted at the beginning of Sec.~\ref{sec:review} for the case of $f_1(Q^2)=1$, $f_2(Q^2)=0$.
This equation holds for all Lorentz components. In particular, we can solve this expression for $f_B$ for the case $\mu=0$,
\begin{align}
\label{eq:ffb_v1}
f_B(Q^2)=g^2\left(\frac{\Actwo^{0}(P_f,P_i)}{E_f+E_i}
+\Gc(s_B,Q^2, s_B)-\frac{2 \Gc^{0}(P_f,P_i)}{E_f+E_i}
\right).
\end{align}

In Sec.~\ref{sec:results}, we will use the finite-volume matrix elements to first determine $\Ac^0$ and then use Eq.~\eqref{eq:ffb_v1} to obtain the form factor. We will do this for a range of values of the parameter of the theory that we describe in Sec~\ref{sec:DiscPEFT}. By varying this one parameter, we will see the variation of the bound state pole and all of its properties.

\subsection{Non-relativistic limit \label{sec:NR_lim}}
In what follows, we will be considering a non-relativistic QFT. Instead of taking the non-relativistic limit analytically of the formalism above, we will be using the relativistic formalism as is to analyze the results and simply take the non-relativistic limit numerically. We do this by constructing ratios of functions that, in the non-relativistic limit, have a non-zero finite value. Equivalently, we consider quantities, like $q^\star\cot\delta$, that have no explicit dependence on the mass of the particles involved. 

For further clarity, we review the basics of the non-relativistic limit. For single particle states, it is easy to consider this limit by expanding by the momentum to be small relative to the mass, or equivalently, that the mass is large. For example, the single particle dispersion relation can be approximated by
\begin{align}\label{eq:omega}
\omega_{q^\star}&=\sqrt{M^2+q^{\star2}}
=  M +\frac{q^{\star2}}{2M}+\mathcal{O}(q^{\star4}) . 
\end{align}
For two-particle systems, the energy of interest is the non-relativistic energy, which in the CM frame is defined by 
\begin{align}
\label{eq:EstarNR}
E^\star_\NR =E^\star - 2M \,,
\end{align}
with $E^\star =2\omega_{q^\star}$.
If the two particle system carries relative momentum $q^\star< M$, then one see that 
\begin{align}
\label{eq:EstarNR_vs_qstar}
E^\star_\NR  = \frac{q^{\star 2}}{M} +\mathcal{O}(q^{\star 4}).
\end{align} 

In a moving frame, the non-relativistic energy, $E_{\rm NR}$, is still defined as the total energy, $E=\sqrt{E^\star+\mathbf{P}^2}$, minus the energy of the two particles at rest, 
\begin{align}\label{eq:ENRrelations}
E_\NR &=E - 2M 
\nn\\
&= E^\star_\NR + \frac{ \mathbf{P}^2}{4M} +\mathcal{O}(\textbf{P}^4) +\mathcal{O}(\textbf{P}^2 q^{\star 2}),
\end{align}
where in the first term we use the definition of $ E^\star_\NR$, and everywhere else we can approximate $E^\star=2M$ up to corrections that can be ignored at this order.

In the non-relativistic limit, it is common to normalize the states as 
\begin{align}
    \langle \mathbf{P}' \vert \mathbf{P} \rangle_{\rm NR} = \frac{P^0}{M} (2 \pi)^3 \delta^{3} (\mathbf{P} - \mathbf{P}' ),  
\end{align}
which at leading order in the non-relativistic expansion simplifies to, 
\begin{align}
    \langle \mathbf{P}' \vert \mathbf{P} \rangle_{\rm NR} =  (2 \pi)^3 \delta^{3} (\mathbf{P} - \mathbf{P}' ).  
\end{align}
This is the normalization that will be adopted for the non-relativistic theory described in Sec~\ref{sec:DiscPEFT}.

\subsection{Relativistic finite-volume formalism}\label{sec:FV_form}

Having reviewed the desired target observables, we give a brief review of the finite-volume formalism needed to get these scattering amplitudes. We begin with the well-known formalism for relating the finite-volume spectrum of two particles, a discrete set of four-momenta $P_n$, to the purely hadronic scattering amplitude, $\Mc$. Assuming the two-particle states exclusively couple to the $\ell=0$ partial wave, the finite volume spectrum satisfies the L\"uscher quantization condition~\cite{Luscher:1991n1, Rummukainen:1995vs, Kim:2005gf},
\begin{equation}\label{eq:QC}
\Mc^{-1}(P^2_n) + F(P_n,L)  =0 \,,
\end{equation}
where $F$ is a known geometric function,
\begin{align}\label{eq:Ffunc} 
F(P,L) & =   \bigg[  \frac{1}{L^{3}} \sum_{\k  } - \int \frac{ \diff^{3}\k}{(2\pi)^{3} } \bigg] \frac{D\big((P-k)^2\big) }{2\omega_{\k} }\bigg|_{k^0 = \omega_{\mathbf{k}}} \,, %
\end{align}
where $\omega_{\k}=\sqrt{M^2+\k^2}$, analogous to the definition of Eq.~\eqref{eq:omega}.

Similarly, $\2+\Jc^\mu\to\2$ amplitudes can be constrained from finite-volume matrix elements~\cite{Briceno:2015tza,Baroni:2018iau}. In particular, one can show that the finite-volume matrix element of a conserved local vector current $\Jc^\mu$, satisfies 
\begin{equation}\label{eq:BH_form}
L^{3} \bra{P_{f}} \Jc^{\mu}(0) \ket{P_{i}}_L =  \sqrt{ \Rc(P_{f},L) \Rc(P_{i},L) }
\, \Wc_{L,\df}^{\mu}(P_{f},P_{i},L) ,
\end{equation}
where the finite-volume states are normalized to unity, i.e.,
\begin{align}
    \bra{P_f } {{P_i}\rangle}_{L} =  \, \delta_{\textbf{P}_f,  \textbf{P}_i}
\delta_{{P}^0_f,  {P}_i^0}\,.
\label{eq:FV_basis}
\end{align}
 The multiplicative factor $\Rc$ is commonly known as the Lellouch-L\"uscher factor~\cite{Lellouch:2000pv}. Here we write this factor in a form that more closely follows Ref.~\cite{Briceno:2014uqa},
\begin{align}\label{eq:Rdef}
\Rc(P_n,L) 
& = 
 -  \Mc^{-2}(P_n^2)   \left[ \frac{\partial}{\partial P^0} \left( F(P,L) + \Mc^{-1}(P^2)   \right) \right]^{-1}_{P = P_n} \,.
\end{align}

Finally, the rightmost quantity in Eq.~\eqref{eq:BH_form} depends explicitly on the object of interest, namely $\Wc^{\mu}_{\df}$,
\begin{align}
\label{eq:WLdf_vector}
\Wc^{\mu}_{L,\df}(P_f,P_i , L) & = \Wc^{\mu}_{\df }(P_f,P_i) 
+      \Mc(P_f^2)  \Big [  (P_f + P_i)^{\mu}  G(P_f,P_i,L)  - 2   G^{\mu}(P_f,P_i,L)    \Big ]  \Mc (P_i^2) \, ,
 \end{align}
where $G$ and $G^{\mu}$ are the finite-volume analogues of the triangle functions $\Gc$ and $\Gc^\mu$, given by Eqs.~\eqref{eq:Gcal} and \eqref{eq:Gcalmu}, respectively. These functions provide additive finite-volume corrections,
\begin{align}
\label{eq:FV_G}
G  (P_f,P_i,L) &= \bigg[  \frac{1}{L^{3}} \sum_{\k  } - \int \frac{ \diff^{3}\k}{(2\pi)^{3} } \bigg]\frac{ D\big((P_f-k)^2\big)\,D\big((P_i-k)^2\big) }{2\omega_{\k}  } \Big\rvert_{k^{0} = \omega_{\k}},
%%%%%%
\\
\label{eq:FV_Gmu}
G^{\mu } (P_f,P_i,L) &= \bigg[  \frac{1}{L^{3}} \sum_{\k  } - \int \frac{ \diff^{3}\k}{(2\pi)^{3} } \bigg] k^{\mu} \frac{ \, D\big((P_f-k)^2\big)\,D\big((P_i-k)^2\big) }{2\omega_{\k}  } \Big\rvert_{k^{0} = \omega_{\k}}.
\end{align}
One notable difference to their infinite-volume counterparts is that the integral over the $k^0$ component has been done explicitly, and it has been shown that only the $k^{0} = \omega_{\k}$ pole leads to power-law finite-volume effects, with the other poles leading to contributions exponentially suppressed in the finite volume.

The $G$ and $G^\mu$ functions involve sums and integrals over poles. The integrals can, in general, lead to anomalous threshold singularities of the kind given by Eqs.~\eqref{eq:Gcal} and \eqref{eq:Gcalmu}. As a result, the evaluation of the three-dimensional integral using standard numerical techniques can result in slow convergence. To circumvent this, Ref.~\cite{Baroni:2018iau} proposed writing the three-dimensional integral in terms of the standard covariant four-dimensional integral, plus a non-singular correction that can be evaluated with faster numerical convergence. For further details of the formalism and its implementation, we point the reader to Ref.~\cite{Baroni:2018iau}. 

\section{Discretized Pionless EFT for two-nucleon systems}
\label{sec:DiscPEFT}

Here we briefly review the pionless EFT ($\text{EFT}_{\slashed \pi}$)~\cite{Kaplan_1998_1, Kaplan_1998_2} used in this study, presenting specific aspects relevant for this work. This $\text{EFT}_{\slashed \pi}$ will be regularized by placing it in a finite-volume lattice, and the infinite volume properties will be extracted following the formalism described above.

The $\text{EFT}_{\slashed \pi}$ describes non-relativistic, point-like nucleons interacting via contact interactions where the $\pi$ (and all heavier, purely virtual) fields have been integrated out. It is expected to apply to systems of nucleons at energies less than the pion mass.
We apply the theory to a system $\psi = (p(x), n(x))$, represented by a degenerate pair of two-component, non-relativistic spinor fields with mass $M$.  The leading-order hadronic Lagrangian density is given by 
\begin{equation}\label{eq:lag}
    \Lc = \psi^\dagger \left( i \partial_t + \frac{\nabla^2}{2M}\right) \psi + \frac{g_0}{4} ( \psi^\dagger \psi)^2 
\end{equation}
where $\nabla^2$ is the Laplacian operator, $\partial_t$ is the partial derivative with respect to time, $g_0$ is a bare coupling, and $\psi$ is the non-relativistic field for the particles being considered. Higher-order terms include derivatives of the fields and consequently become irrelevant near threshold, which is the kinematic region of interest in this work. The coupling, $g_0$, can be tuned to reproduce a particular infinite volume scattering length of the $\ell=0$ partial wave. Because the effective range and higher-order ERE coefficients are not explicitly tuned, they will be of the order of the lattice spacing to the appropriate power.

Bearing in mind that we are interested in evaluating the form factor of a two-particle state, we could, in principle, extend the Lagrangian to include external vector currents. Instead, we proceed as in Ref.~\cite{Kaplan_1999} and determine the consequences of such currents by evaluating their matrix elements. The current we are interested in is the vector current, $\Jc^\mu(x)$, whose matrix elements are defined using the non-relativistic basis and are given by~\cite{Kaplan_1999} 
\begin{align}
    \langle p(\mathbf{P}') \vert \Jc^\mu(0) \vert p (\mathbf{P}) \rangle_{\rm NR} &=  \frac{(P' + P)^\mu}{2 M} ,
    \\
    \langle n(\mathbf{P}') \vert \Jc^\mu(0) \vert n (\mathbf{P}) \rangle_{\rm NR} &=  0,
\end{align}
where the $n$ states are neutral and the $p$ states have a charge normalized to $1$. Throughout the text, we will be referring to these as the neutron and proton, respectively.
We keep the leading term in the non-relativistic expansion and consider the $\mu=0$ component of the current so that this simplifies further to
\begin{align}\label{eq:NRJ0mel}
    \langle p(\mathbf{P}') \vert \Jc^0(0) \vert p (\mathbf{P}) \rangle_{\rm NR} &=  1,
    \\
    \langle n(\mathbf{P}') \vert \Jc^0(0) \vert n (\mathbf{P}) \rangle_{\rm NR} &=  0.
\end{align}
These are the single-particle matrix elements. The matrix elements for two-particle states will follow from these definitions. Before discussing this, we turn our attention to the implementation of this theory in a finite, discretized volume, in order to determine the spectrum.

We consider the discretized version of this EFT presented in~\cite{PhysRevA.84.043644}.
In this work, the Euclidean transfer matrix for the two-particle system, $\mathcal{T} = e^{-H}$, is calculated in the basis of two discretized, non-interacting, single-particle momentum states. Each of these single particle states carries a three-momentum $\mathbf{p} = \frac{2 \pi}{L}(n_x,n_y,n_z)$, where $n_j \in [ -L/2, L/2)$ for a periodic spatial lattice (assuming even $L$).
With boosts implemented, we enforce periodic boundary conditions in the first Brillouin zone (BZ) $\mathbf{p}_j \in [ -\pi, \pi)$.
For convenience, we work in lattice units, where the spatial lattice spacing is equal to 1. 

Although the particles are distinguished by their interaction with the external current, within the $\text{EFT}_{\slashed\pi}$ they behave as identical fermions with an overall anti-symmetric wavefunction. For this calculation, we focus on the case of a two-nucleon system in the isovector channel. In particular, the proton-neutron state appears with the third component of isospin equal to zero. The total spin of the two-nucleon system is chosen to be zero, which, with the isovector condition, ensures that the transfer matrix receives contributions from the interaction in the $\ell=0$ partial wave. This also leads to a state with total angular momentum equal to zero.
The transfer matrix derived from the Lagrangian of Eq.~\eqref{eq:lag} is diagonal in both isospin and spin spaces since these are conserved quantities in this theory.
Therefore, within a given spin-isospin channel, we only need to keep track of the momentum dependence of the transfer matrix.
In the non-interacting basis described above the transfer matrix is given by
\begin{align}
\label{eq:transexplicit}
\langle \mathbf{p}' \mathbf{q}' | \mathcal{T} | \mathbf{p}\mathbf{q}\rangle_L &= 
\frac{\delta_{\mathbf{p}\mathbf{p}'}\delta_{\mathbf{q}\mathbf{q}'}+\frac{g_0}{L^3}\delta_{\mathbf{p}+\mathbf{q},\mathbf{p}'+\mathbf{q}'}}{\sqrt{\xi(\mathbf{p})\xi(\mathbf{q})\xi(\mathbf{q}')\xi(\mathbf{p}')}} \ ,
\end{align}
where the non-interacting finite volume states are normalized to 1, and we have
\begin{equation}
  \xi(\mathbf{q}) \equiv 1+\frac{\Delta(\mathbf{q})}{M} \ , \ \Delta(\mathbf{q}) \equiv -\frac{1}{2} \langle \mathbf{q}| \nabla_L^2 | \mathbf{q} \rangle_L \ .
\end{equation}
which are obtained from the single-particle free propagator. 
The Laplacian operator, $\nabla_L^2$, is constructed to reproduce the continuum dispersion relation in the kinetic term of the Lagrangian up to periodicity within the Brillouin zone,
\begin{equation}
    \Delta (\mathbf q ) = M ( e^{\frac{\mathbf q ^2}{2 M} } - 1) \ ,  \: \: \mathbf{q}_j \in [ -\pi, \pi)
\end{equation}

Note that in the evaluation of the transfer operator, it is useful to project to the two-particle subspace of total momentum $\mathbf{P}$. This reduces the dimensionality of the total calculation by a factor of $L^3$, and singles out systems with a given total lattice momentum, in which the two particle momentum states $\ket{\mathbf{p}\mathbf{q}}_L$ have to satisfy the relation $\textbf{P}= \textbf{p} +\textbf{q}$.

The two-fermion transfer matrix may be diagonalized analytically in a finite volume to match the coupling constants, $g_0$, to energy eigenvalues. For a zero total momentum system, the non-relativistic energy eigenvalues, $\lambda = e^{-E_{\text{NR}}}$, of $\Tc$ are given by solutions to the equation~\cite{Nicholson_2017, PhysRevA.84.043644},
\begin{equation}
\label{eq:inteq}
    \frac{M}{4 \pi} \frac{1}{c} = \frac{1}{L^3} \sum_{\textbf{p} \in \textrm{BZ}} \frac{1}{e^{-E_{\text{NR}} + \mathbf{p}^2 /M}-1}.
\end{equation}
where we have defined $c \equiv M g_0/(4\pi)$. This equation admits a single negative energy state solution, $E_{\text{NR}}<0$, for any value of $c > 0$ in a finite volume.
With this prescription we can tune to a value, $c_{\text{uni}}$, an $M$- and $L$-dependent critical value,  where the lowest, negative energy state becomes a scattering state in the infinite volume limit for $0 < c < c_{\text{uni}}$, and a bound state for $c_{\text{uni}} > c$. This tuning procedure allows us to focus on extracting the energy spectrum for bound state systems. We vary this coupling, allowing us to consider the possibility where the bound state goes from being deeply bound to a shallow bound state. 

Since we are only interested in the interaction in the $s$-wave state, we also need to project to the relevant cubic irreducible representation. For further details, we refer the reader to App.~\ref{app:proj}.
In practice, we project and diagonalize numerically the transfer matrix of Eq.~\eqref{eq:transexplicit}, to find its eigenvectors and eigenvalues $\lambda_n$.
The non-relativistic energy spectrum can then be recovered from the eigenvalues of the transfer operator with the relationship $E_{\text{NR},n}=-\log(\lambda_n)$.
The eigenvectors of the transfer matrix are also the eigenvectors of the finite volume Hamiltonian, $\ket{P_n}_L$, with total energy $E_{\text{NR},n} +2M$ and total momentum $\mathbf{P}_n$. We use these eigenvectors in the two-particle momentum basis to calculate the matrix elements as described in the next section.

\subsection{Finite-volume two-nucleon matrix elements}
\label{sec:FVME}

The formalism of Sec.~\ref{sec:FV_form} presents the matrix element with the current in position space. We evaluate matrix elements with the current in momentum space, so we Fourier transform the matrix element, 

\begin{align}
\label{eq:FV_ME_FT}
    L^3 \langle P_f \vert\Jc^\mu (0)\vert P_i \rangle_L = \langle P_f \vert \widetilde\Jc^\mu  (0,\mathbf{P}_f-\mathbf{P}_i)\vert P_i \rangle_L .
\end{align}

In order to evaluate the finite-volume matrix elements, we employ the finite-volume eigenvectors, $\vert P_n \rangle_L$, obtained from the numerical diagonalization of the transfer matrix.
These eigenvectors can be expressed in the basis of the non-interacting two-particle momentum states used to define the transfer matrix
\begin{align}
    \vert P_n \rangle_L = \sum_{\mathbf{p}\in \textrm{BZ}} a_{\mathbf{p}} (P_n) \vert \mathbf{p} (\mathbf{P}_n-\mathbf{p}) \rangle_L\,,
\end{align}
with the eigenvector components $a_{\mathbf{p}} (P_n)$. Note that $(\mathbf{P}_n-\mathbf{p})$ takes the place of the now-eliminated second degree of freedom, $\mathbf{q}$, in the basis; this is because every nonzero component of the state must have total momentum $\mathbf{P}_n$. 

We recall that the electromagnetic current will couple only to the proton in our system, so there is no single-particle matrix element for the neutron. The discretized EFT requires single-particle form factors for the charged particles in the system, which will then scale the contributions of the electromagnetic currents on the two-particle states.
The single-particle form factor for the structureless proton used in our model is simply equivalent to its charge, which we recall is normalized to 1 for this system. We can write the proton matrix element of the electromagnetic current carrying momentum $\mathbf{k}$ as 
\begin{align}
    \langle \mathbf{p}' \vert \widetilde\Jc^0(0, \mathbf{k}) \vert \mathbf{p} \rangle_L =  \frac{\omega_{\p'} + \omega_{\p}}{2 M} \delta_{\p', \p + \k} \approx \delta_{\p', \p + \k} \,,
\end{align}
where in the last step we assume a non-relativistic limit, similar to what is done for Eq.~\eqref{eq:NRJ0mel}.

We use the single-particle matrix element to calculate the matrix element of the current in the non-interacting two-particle momentum basis. Similar to the transfer matrix, we do not need to keep track of the isospin part of the wavefunction since the electromagnetic current is diagonal in flavor space.
We also do not need to keep track of the spin part when the initial and final states have spin projection $S_z=0$. In this case, the proton and neutron spins are anti-correlated, and the proton spin-flip matrix elements cannot contribute to the two-particle matrix element, effectively making the matrix element diagonal in spin space.
This also means that we can calculate the current matrix element in the two-particle non-interacting basis, assuming the first particle is the proton. Doing so, we find,
\begin{equation}
    \langle \mathbf{p}' \mathbf{q}' | \widetilde\Jc^0(\mathbf{k}) | \mathbf{p} \mathbf{q} \rangle_L
    = \delta_{\p', \p + \k}\,\delta_{\q', \q}\,,
\end{equation}
Finally, the matrix element of the eigenstates of the Hamiltonian can be derived in terms of the components $a_\p(P_n)$,
\begin{equation}
   \langle P_f \vert \widetilde\Jc^\mu  (0,\mathbf{k})\vert P_i \rangle_L 
 =\delta_{\P_f, \P_i + \k}\sum_{\p\in \textrm{BZ}}a_{\p}^*(P_f) a_{\p-\k}(P_i).
\end{equation}

We proceed to analyze the spectrum and matrix elements obtained from the discretized EFT following the formalism described in Sec.~\ref{sec:FV_form} to extract the infinite volume bound state form factor $f_B$.
In particular, we explore the variation of the form factor with the coupling $c$, and we pay particular attention to finite volume effects on the determination of $f_B$.

\begin{figure}[t]
    \centering
    \includegraphics[width=0.9\linewidth]{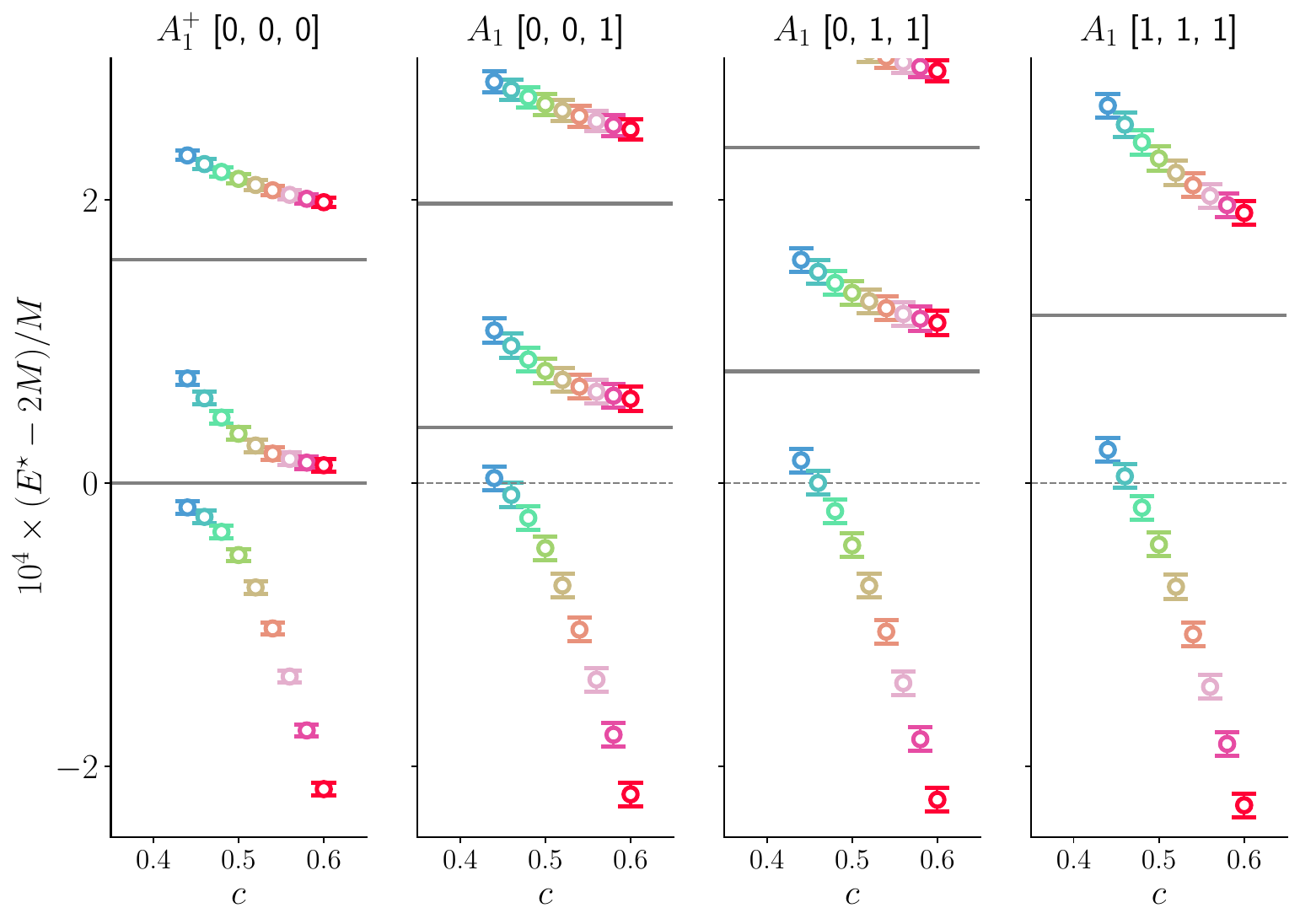}
     \caption{Shown is the finite-volume spectra for the $A_1^{(+)}$ cubic irreps for a range of values of $c$ and fixed values of $L = 10$ and $M = 50$. The errors are introduced as discussed in the main body of text. The solid lines show the non-interacting spectrum, and the dashed line shows the two-particle threshold.
    }
    \label{fig:spectra1}
\end{figure}

%%%%%%%%%%%%%%%%%%%%%%%%%%%%%%%%%%%%%%%%%%%%%%%%%%%%%%%%%%%%%%%%%%%%%%%%%%%%%%%%%%%%
%%%%%%%%%%%%%%%%%%%%%%%%%%%%%%%%%%%%%%%%%%%%%%%%%%%%%%%%%%%%%%%%%%%%%%%%%%%%%%%%%%%%
\section{Analysis of finite-volume spectra and matrix elements}
\label{sec:results}

In this section, we present the main results of this work. Following the steps outlined in the previous section, we determine the finite-volume spectrum of two particles for values of $c$ ranging from $0.44$ to $0.60$. In all scenarios, the theory supports a bound state. For the smaller values of $c$, the bound state is shallow, while for the larger values of $c$, the bound state becomes deeply bound. The expectation is that for the deep bound state, the finite-volume formalism should make little systematic difference, with the leading finite volume effects scaling like $\mathcal{O}(e^{-\kappa L})$ for bound states~\cite{Davoudi_2011, Briceno:2020xxs}, while for a shallow bound state, these corrections are of order $\mathcal{O}(1)$, and the formalism is needed to remove these finite-volume artifacts.

Because we are interested in the non-relativistic limit of the formalism reviewed in Sec.~\ref{sec:FV_form}, we will take the mass $M$ to be large compared to the momenta considered. We find that for $M =50$ in lattice units, we have sufficiently recovered the non-relativistic limit. In what follows, we fix this parameter to this value. 

\subsection{Finite-volume spectrum and purely hadronic observables}
\label{sec:FVS}

Given a fixed value of $\textbf{P}=\frac{2\pi}{L}\textbf{d}$, with $\textbf{d}$ being a triplet of integers, we determine the two-particle finite-volume spectra. The results are shown in Fig.~\ref{fig:spectra1} for the range of values of $c$ considered. We only report the spectra for the $A_1$ irreps up to boost vector of $\textbf{d} = [1,1,1]$. We compare this with the non-interacting spectrum, which is given by 
\begin{equation}
\label{eq:free}
    E_{\text{free}} = \sqrt{ \left(\frac{2 \pi}{L} \mathbf{n}_1\right)^2 + M^2 } +  \sqrt{ \left(\frac{2 \pi}{L} \mathbf{n}_2\right)^2 + M^2 } , \hspace{.2in} E^*_{\text{free}} = \sqrt{ E_{\text{free}}^2 -\textbf{P}^2},
\end{equation}
such that $\textbf{n}_1 + \textbf{n}_2 = \textbf{d}$.  The free energy levels are plotted as gray bands, and the dashed line is the two-particle threshold. 

\begin{figure}[t]
    \centering
    \includegraphics[width=\linewidth,trim={0.2cm 0cm 0cm 0cm}, clip]{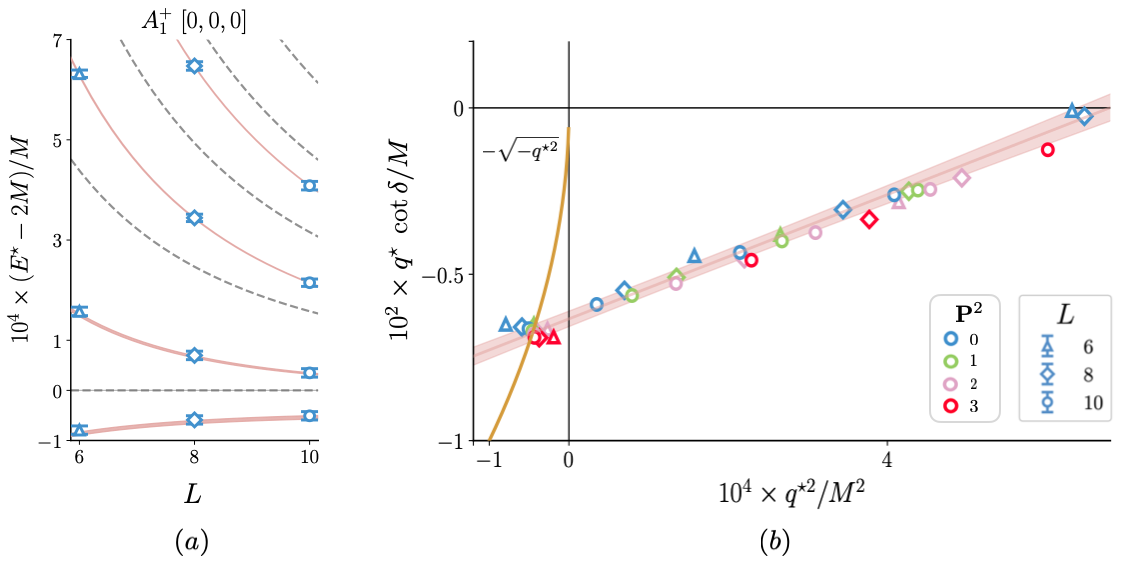}
     \caption{Shown is the procedure for constraining the scattering amplitude from the finite-volume spectrum using the L\"{u}scher quantization condition in Eq.\ref{eq:QC} for $c=0.5$.
     Using the finite-volume energy levels in Fig.~\ref{fig:spectra1}, the ERE parametrization in Eq.~\eqref{eq:ERE} is numerically constrained. In (a), we show the result of inverting this procedure, allowing us to predict the finite-volume spectrum as a function of the lattice size in any frame, where we present the case of the rest frame. The red line is the predicted energy levels with error, while the points are the obtained finite-volume spectrum, and the dashed grey lines are the non-interacting energy levels. In (b), we show the resulting $1 \sigma$ band for $q^\star \cot \delta / M$ alongside the finite-volume spectrum and the bound state condition, $-\sqrt{-q^{\star 2}}$.
     }
    \label{fig:spectra_fit}
\end{figure}

Although the diagonalization of the transfer matrix is exact, we introduce synthetic random uncertainties to study error propagation.~\footnote{The introduction of artificial errors into continuum $\pi$-less EFT was recently explored by introducing Gaussian fluctuations to quantities that are expected to be extracted from LQCD, namely the FV energy eigenvalues and three-point functions~\cite{Davoudi_2022}.} In this analysis, synthetic uncertainties on the four lowest-lying energies in each moving frame and for each coupling are introduced through randomly-generated distributions on the $q^{\star2}$, defined in Eq.~\eqref{eq:EstarNR_vs_qstar}, extracted from each corresponding energy from the transfer matrix. These distributions of momenta have central values equal to $q_i^{\star2}$ and absolute errors, $\sigma_{i}$. We describe below how we generate these errors.  

First, we recognize that the spectra with $\textbf{d}=[0,0,0]$ have smaller discretization errors. Using the unboosted spectrum solely, we do an uncorrelated fit with uniform errors to the quantization condition, Eq.~\eqref{eq:QC}, applying the effective range parameters of Eq.~\eqref{eq:ERE} to describe the scattering amplitude. Then, we use the quantization condition to generate a model spectrum ($E_{\rm mod.}$) assuming these values of the effective range parameters, including for every boost considered. For the resulting spectrum, the systematic error quoted is $\sigma_{i} = |E_{\rm mod.,i}-E_{i}|$. Additionally, we introduce a correlation of $50\%$ across all energy levels generated within a fixed volume. For energy levels from different volumes, we keep them uncorrelated. This explains the uncertainties appearing in the energy levels in Fig.~\ref{fig:spectra1}.

Given these uncertainties and correlations for the spectra, we fit the spectrum using the quantization condition, Eq.~\eqref{eq:QC}. We perform a simultaneous fit of the 4 lowest energies in each frame with $\textbf{d}^2 \leq 3 $ for three different lattice sizes. An example showing this procedure is shown in Fig.~\ref{fig:spectra_fit} for $c=0.5$. Figure~\ref{fig:spectra_fit}($a$) shows the spectrum obtained for this coupling, which has been parametrized and numerically constrained. The resulting fit is shown as a continuous curve as a function of volume. Figure~\ref{fig:spectra_fit}($b$) shows the resulting $1\sigma$ band for $q^\star\cot\delta / M$ as a function of $q^{\star 2} /M^2$, alongside the energy levels used to constrain the quantization condition. Also shown is the $\sqrt{-q^{\star 2}}/M$ curve below threshold. We see a clear crossing of these curves, which, from Eq.~\eqref{eq:BS_pole}, one sees is clear evidence of a bound state pole in $\Mc$ below threshold.

\begin{figure}[t]
    \centering
    \includegraphics[width=\linewidth]{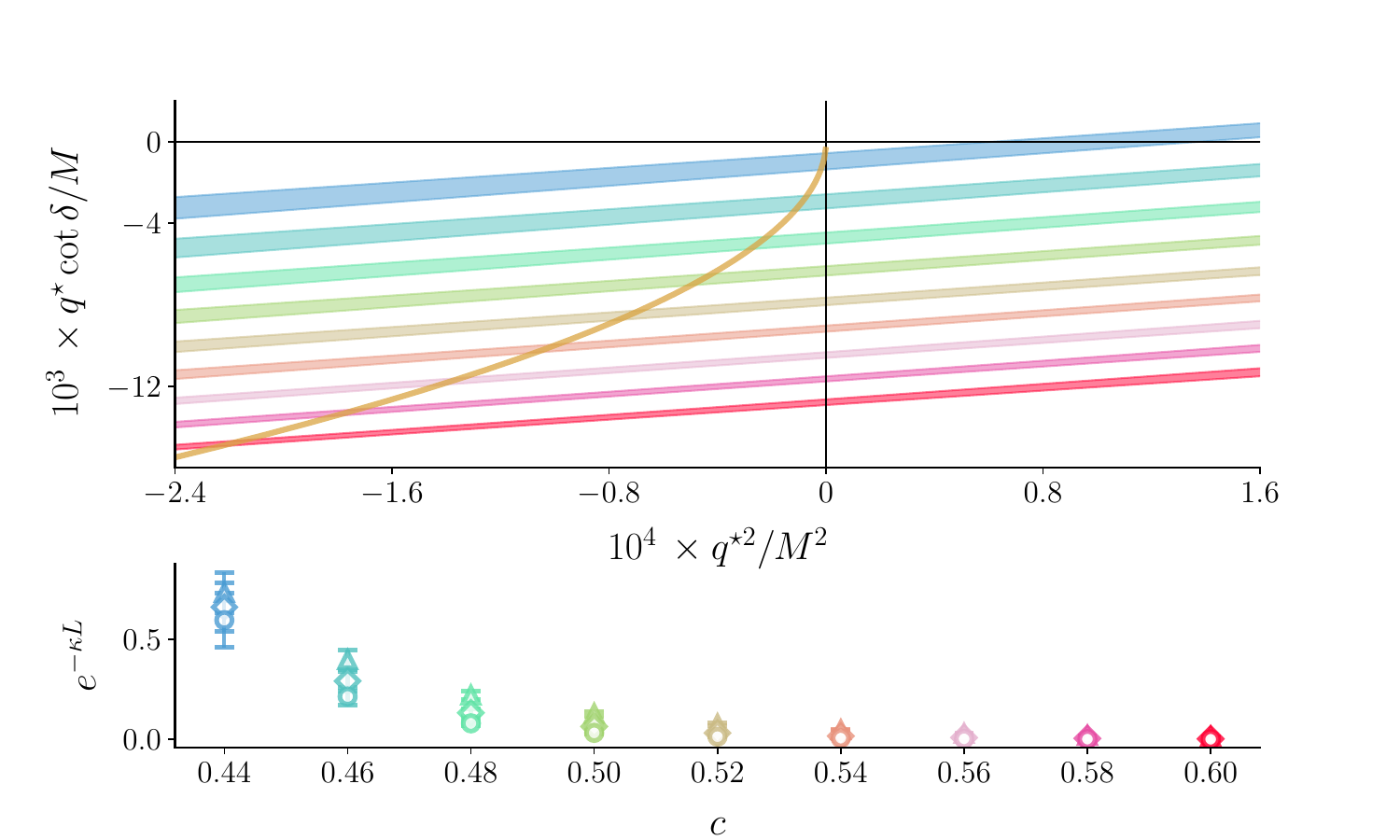}
    \caption{In the top panel, we show the resulting parametrized $q^\star \cot \delta / M$ for the different values of the coupling $c$ considered in Fig.~\ref{fig:spectra1} as a function of the COM momentum compared to the bound state condition, as given in Eq.~\eqref{eq:BS_pole}. In the bottom panel, we show the expected finite-volume corrections on the matrix elements as a function of the coupling of the theory.}
    \label{fig:masterphase}
\end{figure}
 
We repeat this exercise for a range of values of the couplings $c = [0.44,0.6]$. The resulting fitted $q^\star \cot\delta/M$ with uncertainties for these parameters are given in the top panel of Fig.~\ref{fig:masterphase}. We see that for the smallest value of $c$, one obtains a remarkably shallow bound state, while for the largest values of $c$, one obtains a rather deep bound state. As the state becomes shallower, its size grows inversely with the binding momentum. As a result, not only the energy of this state, but also its matrix elements, will be more sensitive to the volume. We also calculate the expected size of the leading-order finite-volume corrections derived in Refs.~\cite{Davoudi_2011, Briceno:2020xxs}, where the finite-volume effects corrected by the formalism are of order $\exp({- \kappa L})$, in the bottom panel of Fig.~\ref{fig:masterphase} obtained using the binding momentum found with Eq.~\eqref{eq:BS_pole}. We will examine the effect of the formalism on these artifacts for the matrix elements in the subsequent sections.

Before presenting the matrix elements and their analysis, we first evaluate the remaining ingredient required for the analysis, specifically the Lellouch-L\"uscher factors. These were defined in Eq.~\eqref{eq:Rdef} with the symbol $\Rc$. This definition of the Lellouch-L\"uscher factors results in a complex number, but the phases are given directly from the scattering phase shift. It is easy to verify that the following combination is purely real, 
\begin{equation}
    \widetilde{\Rc}( P_n, L) = \Mc^2 (s_n (L)) \Rc( P_n, L).
\end{equation}

One important check of this quantity was given in Ref.~\cite{Briceno:2019nns}, where it was shown that for a bound state in the large $L$ limit, $ \widetilde{\Rc}$ is equal to,
\begin{align}
     \widetilde{\Rc} (P_n, L)\bigg|_{P_n^0\approx E_B}
     &= \left[ \frac{2 E_B}{g^2} + \Oc(e^{-\kappa L}) \right ]^{-1}
     \\
     &= \frac{g^2}{2 E_B} + \Oc(e^{-\kappa L}) ,
     \label{eq:R_bs_lim}
\end{align}
where $g$ is given by Eq.~\eqref{eq:g2}, and $E_B = \sqrt{s_B +\textbf{P}^2}$. The factor of $2E_B$ corrects for the fact that the finite-volume states are normalized to $1$, and the coupling $g$ can be understood as the infinite-volume wavefunction renormalization of the bound state. This is to say that this combination of factors conspires to relate the finite-volume matrix elements of a bound state to its infinite-volume matrix elements. 

\begin{figure}[t]
    \centering
    \includegraphics[width=0.95\linewidth]{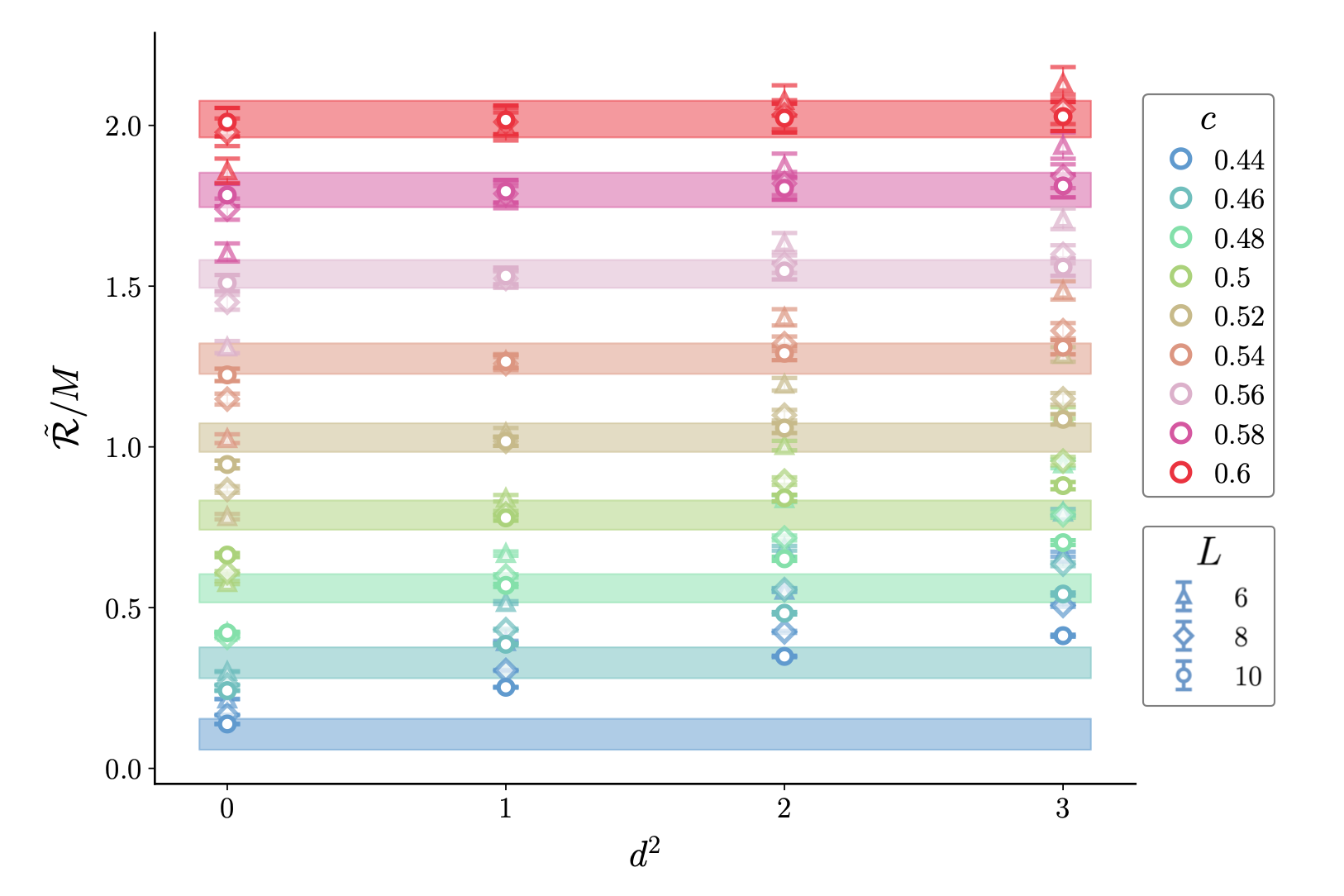}
    \caption{Shown are the values of $2E_n\,\widetilde{\Rc}= 2E_n\,\Mc^2\, \Rc$ for the ground state of all volumes and values of the coupling. The errors are due to the propagation of errors of the fitted effective range parameters. The straight lines are the infinite-volume prediction of $2E_n\,\widetilde{\Rc}$, which is equal to $g^2$ .}
    \label{fig:LLfactor}
\end{figure}

More importantly, within the context of this work, this gives us an analytic expression to compare our numerical evaluations of $\widetilde{R}$. In Fig.~\ref{fig:LLfactor}, we plot the product $2 E_n\widetilde{\Rc}$ for the ground state of all of our volumes, boosts, and couplings. For comparison, we show the value of $g^2$, which is what Eq.~\eqref{eq:R_bs_lim} predicts this value should approach. We see that, as one would expect, these agree moderately well for the deep bound states, but there remain sizable deviations for the shallower bound states.  

\subsection{Finite-volume matrix elements and \texorpdfstring{$f_B$}{fB}}

Having all the ingredients for the analysis of the matrix elements, we proceed to the results for the matrix elements. Although we have evaluated all matrix elements for the two-particle sector, here we only report the matrix elements for the transitions with the initial state being the ground state with $\mathbf{P}=0$, and the final state corresponding to the ground state across the different boosts. We report the matrix elements for the three volumes considered in the spectrum analysis. As discussed in Sec.~\ref{sec:FVME}, we evaluate the finite-volume matrix elements of the Fourier transformed matrix currents, see Eq.~\eqref{eq:FV_ME_FT}. Furthermore, we only quote the $\mu=0$ component of the current. This component has the nice feature that if we assume that finite-volume state is exponentially close to the infinite-volume bound state and take the non-relativistic limit, the finite volume matrix element is exponentially close to the bound state form factor. To show this relation, we start from the left-hand side of Eq.~\eqref{eq:FV_ME_FT}, and use various previously derived relations,
\begin{align}
    \label{eq:FV_ME_FT_v2}
    \langle P_f \vert \widetilde\Jc^{\mu=0}  (0,\mathbf{P}_f-\mathbf{P}_i)\vert P_i \rangle_L
    &=
    \frac{1}{2 \sqrt{E_f\,E_i} } {2 \sqrt{E_f\,E_i} L^3} \langle P_f \vert\Jc^{\mu=0} (0)\vert P_i \rangle_L\,, \nn\\
      %%%%%%%%%%%%%%%%%%%%%%%%%%%%%%%
    &\approx
    \frac{1}{2 \sqrt{E_f\,E_i} }  \langle P_f \vert\Jc^{\mu=0} (0)\vert P_i \rangle_{\mathrm{rel.}}\,, \nn\\
     %%%%%%%%%%%%%%%%%%%%%%%%%%%%%%%
    &=
    \frac{1}{2 \sqrt{E_f\,E_i} }  (E_f+E_i) f_B(Q^2)\,,
    \nn\\
    %%%%%%%%%%%%%%%%%%%%%%%%%%%%%%%
    &\approx
     f_B(Q^2).
    %%%%%%%%%%%%%%%%%%%%%%%%%%%%%%%
\end{align}
In the second equality, we used the fact that for stable particles the proportionality factor between finite-volume states normalized to $1$ and relativistic states is $\sqrt{2 E L^3}$. In the last equality, we used the non-relativistic limit. In other words, if finite-volume effects and relativistic effects are negligible, this matrix element should be a single-valued analytic function of $Q^2$. 
\begin{figure}[t]
    \centering
    \includegraphics[width=0.95\linewidth]{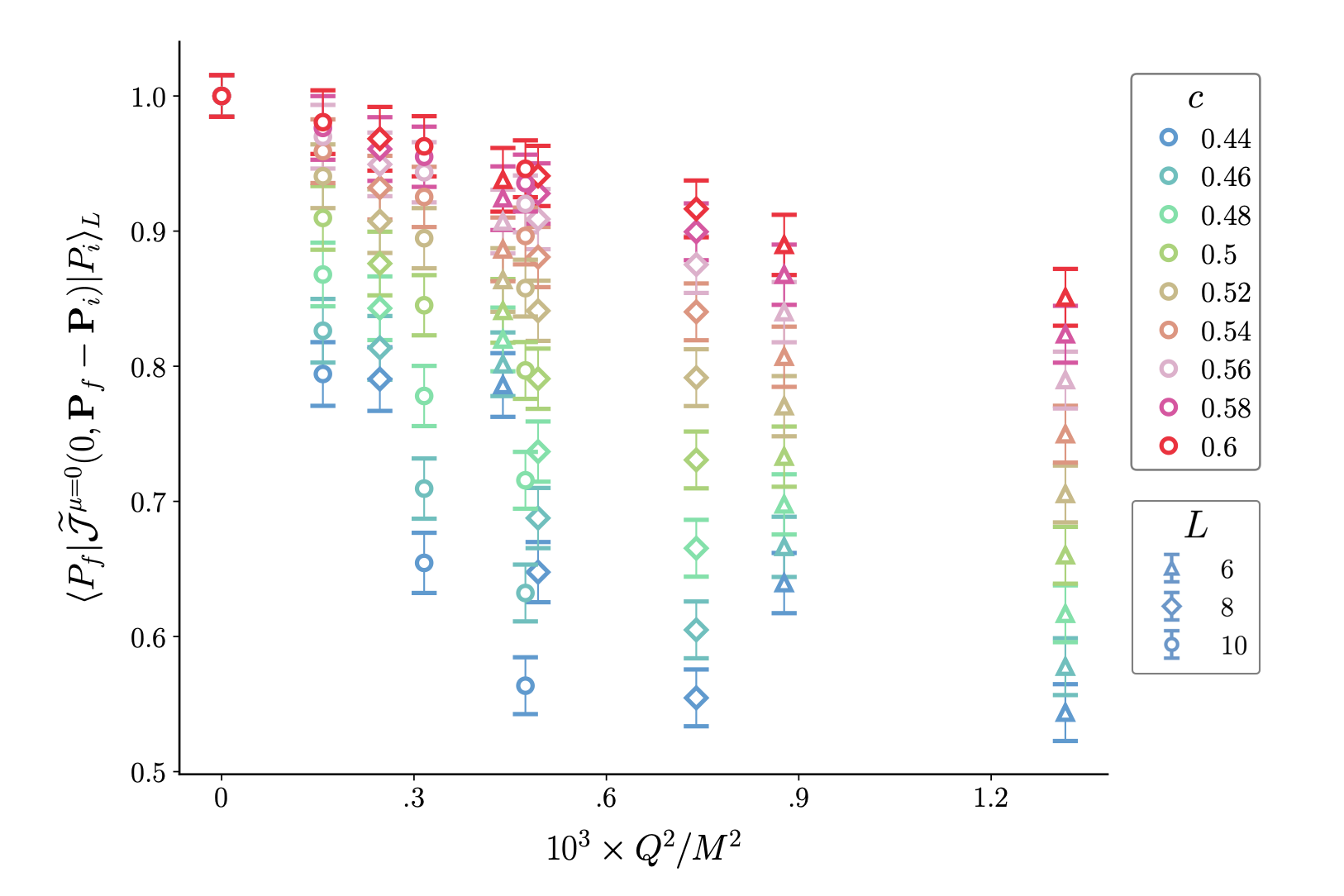}
    \caption{Shown are the finite-volume matrix elements, obtained using the construction in Sec.~\ref{sec:FVME}, for the transition from the ground state to moving frames mediated by the temporal electromagnetic current, interacting with only one of the particles, as a function of $Q^2$. The matrix elements are calculated for the finite-volume spectrum across various values of the coupling and lattice sizes.}
    \label{fig:FVME}
\end{figure}

The finite-volume matrix elements for the ground state are shown in Fig.~\ref{fig:FVME}. 
We add synthetic uncertainties to the matrix elements, $\sigma_{\text{ME}}$, proportional to the synthetic relative error of the CM energies of the initial and final states; these errors are shown in Fig.~\ref{fig:spectra1}.
More specifically, we add the relative errors in quadrature,
\begin{equation}
    \sigma_{\text{ME}}\propto \sqrt{\left(\frac{\sigma_{E^\star_i}}{E^\star_i}\right)^2
+\left(\frac{\sigma_{E^\star_f}}{E^\star_f}\right)^2}\,
\end{equation}
and we adjust the proportionality constant such that the relative error of the matrix elements at $Q^2=0$ is approximately~$1.5\%$, which is the typical precision achieved by modern LQCD calculations of nuclear matrix elements.

As expected, we see that for the deep bound states, the matrix elements qualitatively behave as a form factor with a negative slope of relatively small magnitude. This suggests that it has a small charge radius, i.e.\ it is relatively compact. On the other hand, we see that for the shallowest bound state, the matrix element seems to be multivalued, or it has a large unaccounted-for systematic error. We note that at $Q^2 = 0$, all of the matrix elements are equal to one, as expected by charge conservation, so we only show one matrix element at this value.

\begin{figure}[t]
    \centering
    \includegraphics[width=1\linewidth]{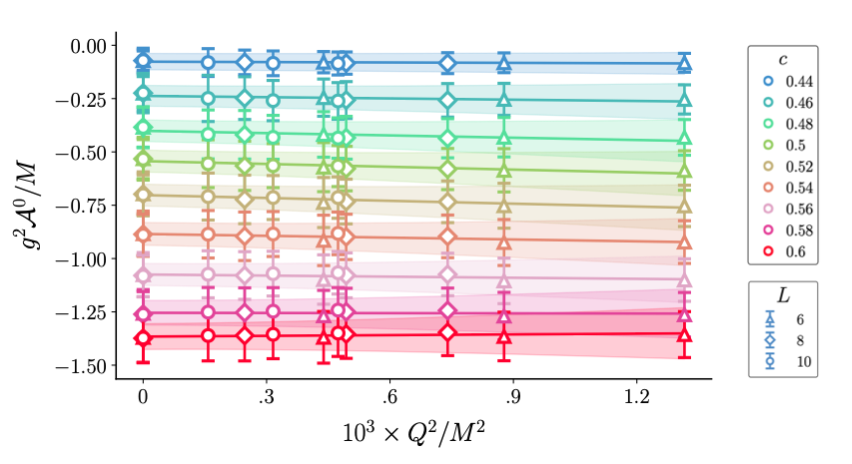}
    \caption{We present the results for the dimensionless quantity $g^2 \Ac^0 /M$, where $g^2$ is the residue of the bound state and $\Ac^0$ is obtained using Eq.~\eqref{eq:Wdf_df}. For each value of $c$, we fit the discrete points using the parametrization given in Eq.~\eqref{eq:poly}.
    }
    \label{fig:a22}
\end{figure}

Given the matrix elements, we can use Eqs.~\eqref{eq:BH_form} and \eqref{eq:WLdf_vector}, to obtain $\Wc_\df^{\mu =0}$. From this, we use Eq.~\eqref{eq:Wdf_df}, to relate $\Wc_\df^{\mu =0}$ to $\Ac^{\mu =0}$. These steps require the evaluation of the previously discussed purely hadronic amplitude $\Mc$, the $\Rc$ factor, as well as the finite- and infinite-volume triangle functions. For the latter, we use the code presented in Ref.~\cite{Baroni:2018iau}. Given these steps, in Fig.~\ref{fig:a22}, we plot $\Ac^{\mu =0}$ as a function of $Q^2$.
In this case, the initial state of the matrix element is at rest, i.e.\ $\mathbf{P}_i=0$, so that all the resulting values of $\Ac^{\mu =0}$ shown in Fig.~\ref{fig:a22} are all evaluated in the frame corresponding to that of a fixed target experiment. This is needed to remove frame dependence, which mixes temporal and spatial components, from the four-vector $\Ac^{\mu}$.

The formalism seems to have done two remarkable things. The first is to have removed all power-law finite-volume effects. This is most dramatic for the shallow bound state, where the $\Ac^{\mu =0}$ function is no longer multi-valued. This is, of course, the intended purpose of the formalism, which provides a check of its effectiveness. The second is that $\Ac^{\mu =0}$ seems to have no dependence on $Q^2$. In other words, the formalism relates finite- and infinite-volume matrix elements up to an overall real-valued function that, at least for this case, is a constant. This suggests that all of the $Q^2$ dependence of these amplitudes, and consequently the bound state form factor, is driven by the triangle functions, or equivalently, by the so-called \emph{anomalous threshold}.

In general, $\Ac^\mu$ is a Lorentz vector, which can be Lorentz decomposed in terms of energy-dependent form factors, $F_i(s_f, Q^2, s_i)$. In general, this might be the best procedure for performing a detailed analysis of this function. However, having only evaluated $\Ac^{\mu=0}$ in a fixed frame, nothing prevents us from parameterizing this quantity directly. Given how structureless this function appears to be, this is what we analyze. 

We consider a total of 8 different types of parametrizations for fitting $\Ac^0$. Given the exploratory nature of this work, we just report the results from a single fit, given by a linear parametrization
\begin{equation}
\label{eq:poly}
     \Ac^0_{\rm fit} ( Q^2 ) = x_0 + x_2 Q^{2}.
\end{equation}
The results from the other 7 parametrizations statistically agree with the result of this fit. 

\begin{figure}[t]
    \centering
    \includegraphics[width=\linewidth]{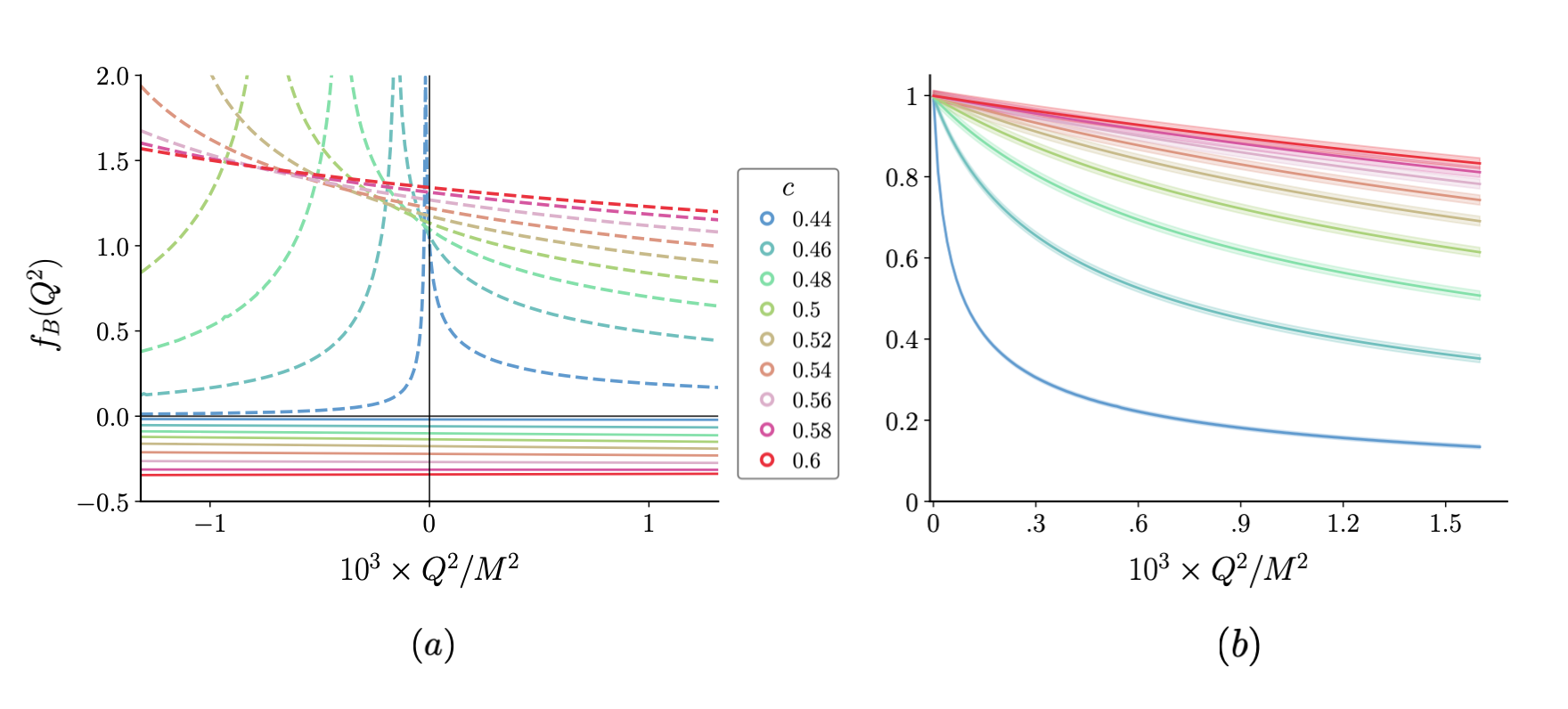}
    \caption{Shown is the final result for the bound state ground state elastic form factor using Eq.~\eqref{eq:ffb} and the parametrized form of $\Ac^0$ in Eq.~\eqref{eq:poly} for a range of couplings from a deep- to shallow-bound states.
    In ($a$), we show the decomposition of factors contributing to the bound state form factor, where the solid lines are the $\Ac^0$ contributions (Eq.~\eqref{eq:ffb_Ac}) and the dashed lines are the contributions from the scalar and vector triangle functions (Eq.~\eqref{eq:ffb_tri}) for a range of couplings. In ($b$), we show the resulting form factor with propagated errors.}
    \label{fig:FF}
\end{figure}

In Fig.~\ref{fig:FF}($a$), we break down the contribution to the form factor, as defined in Eq.~\eqref{eq:ffb}. These two include the $\Ac$-dependent contribution, 
\begin{align}
\label{eq:ffb_Ac}
f_{B,\Ac}(Q^2)\equiv g^2 \frac{\Actwo^{0}(s_B,Q^2, s_B)}{E_f+E_i}.
\end{align}
and the triangular contribution, 
\begin{align}
\label{eq:ffb_tri}
f_{B,\rm tri.}(Q^2)\equiv g^2\left(\Gc(s_B,Q^2, s_B)-\frac{2 \Gc^{0}(s_B,Q^2, s_B)}{E_f+E_i}
\right).
\end{align}
Although the form factor is being constrained in the spacelike region $Q^2\geq 0$, we show these functions not only for $Q^2\geq 0$, but also for timelike values. This allows us to see the role of the anomalous threshold~\cite{PhysRev.111.1187, PhysRev.114.376,Nambu:1958zze,RevModPhys.33.448}, which is the point where the logarithmic singularity in the triangle functions diverges. This divergence occurs at the current virtuality $Q^2=Q_A^2$, where 
\begin{equation}
\label{eq:qa}
    Q_A^2 = -4 s_B  \left( 1 - \frac{s_B}{4 M^2}\right)\,.
\end{equation}
We see that as the bound state becomes shallower, the anomalous threshold approaches the physical region, $Q^2\geq 0$.
Having determined $\Ac^0$, we are finally at a stage to evaluate $f_B$, using Eq.~\eqref{eq:ffb}.

The bound state form factor $f_B$ is shown in Fig.~\ref{fig:FF}($b$), which is the sum of the two contributions shown in Fig.~\ref{fig:FF}($a$). We show the form factors extracted for different values of the coupling $c$, as a function of $Q^2$.
We immediately see that the form factors obtained are smooth, single-valued functions, as one would expect.
The form factor has to be exactly equal to $1$ at $Q^2=0$ by charge conservation. What is more interesting to see is that away from this point, the behavior of the form factor is qualitatively explained by the vicinity of the anomalous threshold to $Q^2=0$. As $Q_A^2$ approaches the physical region, the form factor exhibits a larger variation in $Q^2$.

\begin{figure}[t]
    \centering
    \includegraphics[width=0.95\linewidth]{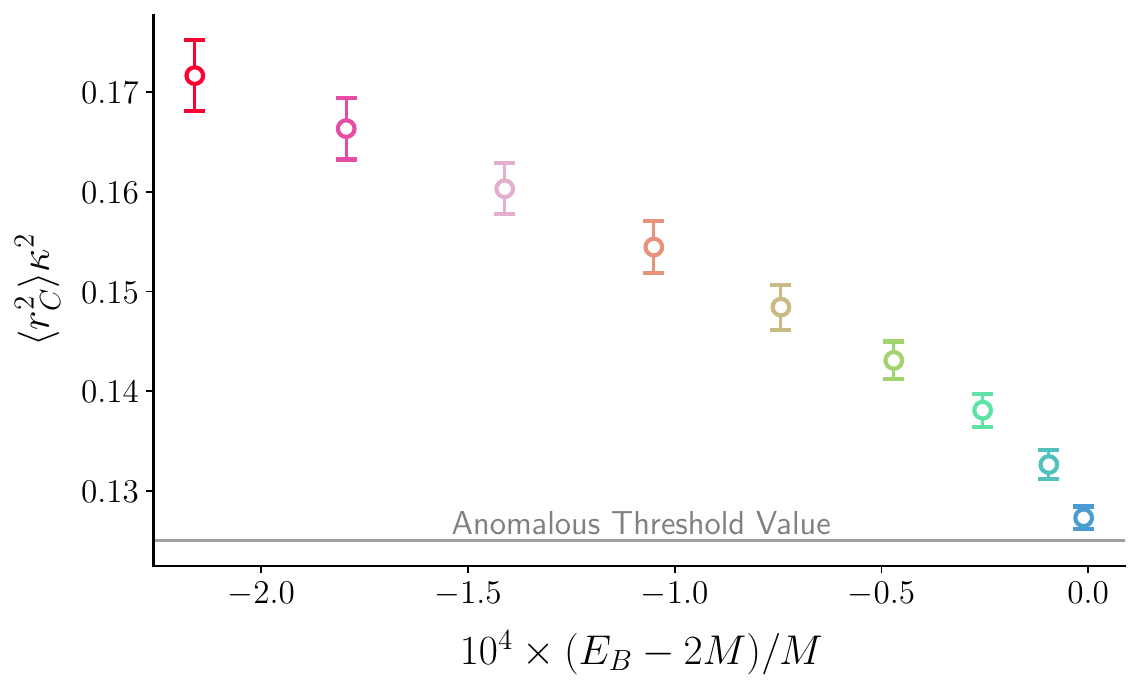}
    \caption{The extracted charged radius of the bound states as a function of the extracted bound state energy compared to the predicted value of the anomalous threshold.}
    \label{fig:radius}
\end{figure}

 Given the form factor, we can now evaluate the charge radius of these bound states. For this, we use the well-known formula relating the radius to the derivative of the form factor as $Q^2=0$, 
\begin{equation}
    \langle r^2_C \rangle = -6 \frac{d f_B (Q^2 ) }{d Q^2} \big \vert_{Q^2 = 0}.
\end{equation}
The resulting radii are shown in Fig.~\ref{fig:radius}. For comparison, we show the leading behavior of the charge radius of a two-particle bound state when the binding momentum goes to zero,
\begin{equation}
\label{eq:rlo}
    \langle r^2_C \rangle  = \frac{1}{8 \kappa^2}.
\end{equation}
This relationship is independent of the details of the interaction, as long as it has a finite range.
Equation~\eqref{eq:rlo} is universal in that it can be derived both from a non-relativistic and a relativistic approach.
In the case of the relativistic formalism, the triangle contribution of Eq.~\eqref{eq:ffb_tri} is directly responsible for this behavior, while the $\Ac$-dependent contribution vanishes in this limit.
In the non-relativistic case, Eq.~\eqref{eq:rlo} can be derived by employing the asymptotic behavior of the wavefunction first derived in Ref.~\cite{10.1098/rspa.1935.0010}.
This result also follows from Eq.~\eqref{eq:LOf}, which was derived for the non-relativistic pionless EFT of Ref.~\cite{Kaplan_1999}.
More details on the derivation of Eq.~\eqref{eq:rlo} are given in App.~\ref{app:radius}.

Fig.~\ref{fig:radius} shows that as the bound state becomes increasingly shallow, one recovers the prediction of Eq.~\eqref{eq:rlo}. This is reassuring, given that, as we discussed above, we find that the form factor is largely driven by the anomalous threshold in this limit. This provides one of the most stringent checks in the formalism reviewed in Sec.~\ref{sec:review}.

\begin{figure}[t]
    \centering
    \includegraphics[width=0.95\linewidth]{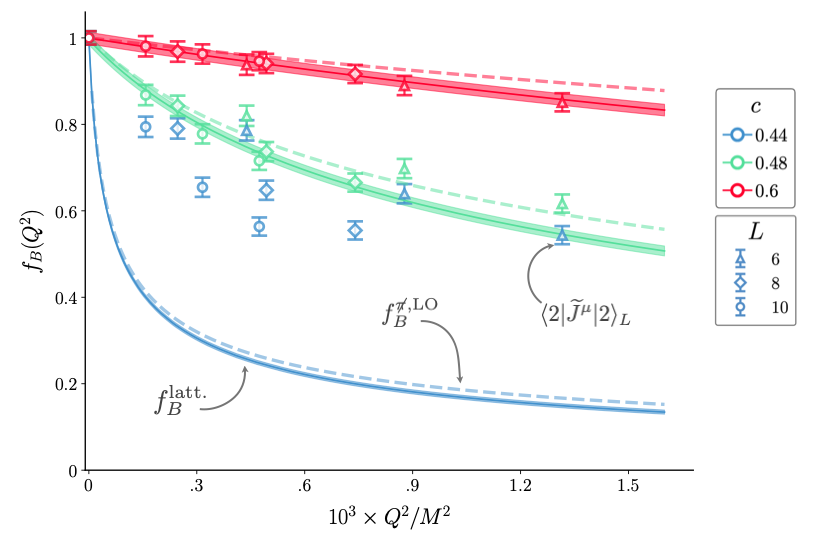}
    \caption{Shown is the final result for the bound state elastic form factor using Eq.~\eqref{eq:ffb} and the parametrized form of $\Ac^0$ in Eq.~\eqref{eq:poly} for three couplings compared to the LO perturbative result from Eq.~\eqref{eq:LOf} and to the finite volume matrix elements defined in Sec.~\ref{sec:FVME}.}
    \label{fig:ff_comp}
\end{figure}

Finally, we review the impact of the overall formalism on the original matrix elements. In Fig.~\ref{fig:ff_comp}, we show the finite volume matrix elements for three values of the coupling compared to the extracted bound state form factors. These three values show clear evidence of the trend described in Ref.~\cite{Briceno:2019nns}. For deep bound states, the finite-volume matrix elements are consistent with the infinite-volume ones. For these cases, the formalism has minimal impact, and most importantly, it does not appear to have any bias with respect to the matrix elements. It can be understood as simply giving a parametrization for the form factor. For the shallower bound states, the formalism gives a form factor that is many standard deviations away from the finite-volume matrix element, removing the large finite-volume artifacts. We also compare the extracted form factors to the ones predicted by LO pionless EFT~\cite{Kaplan_1999}, 
\begin{equation}
\label{eq:LOf}
        f_B^{\text{LO}} (Q^2 ) = \frac{4 \kappa}{Q} \tan^{-1}\left( \frac{Q}{4 \kappa}\right).
\end{equation}
We observe a rather good agreement between this lattice EFT calculation and the analytic leading order result. This could be understood to be a consequence of the fact that the effective ranges obtained for all the couplings are small compared to our range of inverse momenta, explaining why sub-leading effects are suppressed. 

\section{Conclusion}
\label{sec:conclusion}

In this work, we present an exploratory lattice calculation of a two-body matrix element using the LO pionless EFT, which serves as a reasonable toy model for NN scattering at near-threshold energies. As in previous works, e.g., Refs.~\cite{Chandrasekharan:2024iao, Lee_2009,lähde2015nuclearlatticesimulationsusing,Lee_2008,Klein_2015,Alarc_n_2017,Lu_2019,Contessi_2017}, we have determined the finite-volume spectrum of this theory and related it to the infinite-volume scattering amplitude using the well-known L\"uscher formalism. We did this for a range of values of the coupling, allowing for the would-be deuteron to emerge with varying degrees of binding, as one may encounter, e.g., with different pion masses used for LQCD calculations. 

We go beyond previous work by considering the finite-volume matrix elements of this theory for the local conserved vector current. For the deep-bound-state scenario, we see that the finite-volume matrix elements qualitatively resemble the form factor of a small compact object, while for the shallow bound state, the matrix elements exhibit non-monotonic behavior in $Q^2$ when comparing results with different lattice volumes, which is a clear indication of large finite-volume artifacts. By restricting our attention to the matrix elements of the lowest energy levels for each volume, we use the formalism presented in Refs.~\cite{Briceno:2015tza, Baroni:2018iau} to map these matrix elements to the infinite-volume $\mathbf{2} + \Jc^\mu \rightarrow \mathbf{2}$ amplitudes, $\Wc^\mu$, which couple two-particle scattering states via the insertion of a local current. Using the on-shell decomposition of this amplitude presented in Ref.~\cite{Briceno:2020vgp}, we relate this amplitude to a short-distance, purely real function, $\Ac^\mu$. For all the values of the EFT coupling considered, we find that this function can be described as a constant function of $Q^2$. This is a non-trivial observation, given that the resulting form factors have a strong $Q^2$-dependence.  

After performing the full analysis of the matrix elements for each value of the coupling, we observe that all the resulting bound state form factors are monotonic functions of $Q^2$. We also see a clear, smooth behavior in the form factors as a function of the coupling of the theory. For each form factor, we evaluate the charge radius, which is given by the $Q^2$ derivative at $Q^2=0$. We observe that in the limit that the bound state approaches the threshold, our evaluation of the radius asymptotes to the prediction, which assumes that the anomalous threshold completely describes the form factor. 

This study does two major things. First, it provides the best proof-of-principle that the formalism and framework laid out in Refs.~\cite{Briceno:2015tza, Baroni:2018iau, Briceno:2020vgp} for studying $\mathbf{2} + \Jc^\mu \rightarrow \mathbf{2}$ reactions is self-consistent. Second, it explains why this is critical for studying the structural information of shallow bound states, such as the deuteron at near-physical quark masses. 

Going forward, there are two remaining subtleties to be addressed to begin to apply this class of formalism for $NN$ lattice QCD calculations. First, the formalism presented in Refs.~\cite{Briceno:2015tza, Baroni:2018iau, Briceno:2020vgp} is strictly speaking only applicable for spinless particles. For low energies, effects due to the spin of the nucleons are not likely to play an important role, but this must be worked out in detail. Second, we must extend our study to include virtual bound states, which are real-valued poles on the nearest unphysical Riemann sheet of the $NN$ scattering amplitude. The analytic continuation for the purely hadronic amplitude is straightforward. For the $\2+\Jc\to\2 $ amplitude, this analytic continuation is more subtle~\cite{Briceno:2020vgp}. In particular, the prescription proposed in Ref.~\cite{Briceno:2020vgp} is assured to give the correct answer for energies close to the cut in the first Riemann sheet, but it does not provide an unambiguous procedure for continuation away from the cut into the second Riemann sheet.

These two remaining limitations, although technical, can be expected to be removed in the near future. These formal developments, along with the ongoing lattice QCD program, will ultimately allow us to resolve not just the structure of composite states, like the deuteron, but also access a broad class of previously inaccessible electroweak nuclear reactions.

\section*{Acknowledgments}
JM would like to thank Ivan M. Burbano, Marco A. Carrillo, Zohreh Davoudi, Malcolm Lazarow, and Keegan H. Sherman for discussions.
JM is supported in part by the National Science Foundation (NSF) FRHTP program under award No. PHY-2402275, and by the Department
of Physics, Maryland Center for Fundamental Physics, and College of Computer, Mathematical, and Natural Sciences at the University of Maryland, College Park.
JM was supported in part by the U.S. National Science Foundation (NSF) Graduate Research Fellowship Program under Grant
No. DGE-2040435. Any opinions, findings, and conclusions or recommendations expressed in this material are those of the author(s) and do not necessarily reflect the views of the NSF. JM was also supported in part by the U.S. Department of Energy (DOE), Office of Science, Office of Nuclear Physics, under grant contract numbers DE-AC02-05CH11231, the DOE Topical Collaboration “Nuclear Theory for New Physics”, award No. DE-SC0023663, and the U.S. DOE, Office of Science, Office of Workforce Development for Teachers and Scientists, Office of Science Graduate Student Research (SCGSR) program. The SCGSR program is administered by the Oak Ridge Institute for Science and Education (ORISE) for the DOE. ORISE is managed by ORAU under contract number DESC0014664.
FGOG and RAB were partly supported by the U.S. Department of Energy, Office of Science, Office of Nuclear Physics under Award No. DE-SC0025665 and No. DE-AC02-05CH11231. JM and AN were partially supported by the NSF Faculty Early Career Development Program (CAREER) under award PHY-2047185. AWJ acknowledges the support of the USDOE ExoHad Topical Collaboration, contract DE-SC0023598.
\bibliography{bibi} %
\appendix
% %%%%%%%%%%%%%%%%%%%%%%%%%%%%%%%%%%%%%%
\section{Projecting the transfer operator}
\label{app:proj}

In this work, we have projected the transfer matrix to the trivial irrep of the cubic group for the case where $\textbf{d} = [0,0,0]$ and its little groups otherwise. This projection follows directly from previous work~\cite{Luu_2011, G_ckeler_2012, Thomas_2012, Dudek_2012, Drut_2013, Wu_2022}. For completeness, we review some key ideas, beginning with the cubic group. Although we have discretized our volume by a cubic mesh, the reason for needing to project the low-energy spectrum to these irreps is due to the fact that we have also made the volume cubic.

When truncating the space to be finite and cubic, the point group that one is concerned with is the octahedral group $O$ for particles with integer spin. The group contains the proper rotations that transform an octahedron into itself, not including the parity transformation. The group $O$ consists of 24 rotation elements, $R$, which are further categorized into separate conjugacy classes. The group that includes inversion is the $O_h$ group, which is the product of the $O$ group and $C_2 = \{ e, \hat{\sigma }\}$, $O_h = O \otimes C_2$, a group with 48 elements. The two-particle system at rest, $\textbf{d} = [0,0,0]$, will be contained in the irreps of $O_h$, which are labeled as $A_{1}^\pm (1) $, $A_2^{\pm} (1)$, $E^{\pm}(2)$,  $T_1^{\pm}(3)$, and  $T_2^{\pm}(3)$, where the number in parenthesis distinguishes the dimensionality of the group. The character table and rotations for the relevant point groups can be found in Tables.~V and VI of Ref.~\cite{Wu_2022}. 

When considering boosted systems in the CM frame moving with velocity $\mathbf{\beta} = \mathbf{P}/E$, where $\textbf{P} = \frac{2 \pi}{L }\textbf{d}$, the symmetry group further reduces to subgroups of the full octahedral groups, which are referred to as the little groups. In each little group, the allowed transformations are found by considering elements of the full group $S_i$, which obey 
\begin{equation}
    S_i \textbf{d} = \textbf{d},
\end{equation}
leaving only a subset of the total elements of $O_h$ in each moving frame. In constructing each of the rotations and their corresponding conjugacy classes, the allowed momenta in our calculation can have any values of the integer triplet, $\textbf{d}= [n_x, n_y, n_z]$, necessitating the classification of all point groups. 

Once we have all the allowed rotations in each frame, the mapping between the finite-volume irreps and the continuum, $SO(3)$ can be used to identify which irrep contains the partial-wave $\ell=0$. This results in a decomposition of the $\ell$-wave representations of $SO(3)$ in terms of the irreps of the octahedral group, so that different linear combinations of irreps correspond to a wave of $SO(3)$. A thorough treatment of this problem can be found in Ref.~\cite{Luu_2011}, where, for example, for systems at rest only the $A_1^\pm$ representation contributes to $s$-wave scattering but also contains contamination of waves with $\ell=4,\dots$. A similar pattern is observed in moving frames, where a higher degree of partial wave mixing necessitates the identification of certain partial waves. 

To project the transfer matrix to the irreducible representations of the cubic group ($r$), we must construct a projection operator that acts on the momentum states of the transfer operator. The projector operator that we define is 
\begin{align}
\label{eq:projection}
    \langle \mathbf{p} \: \mathbf{q} \vert 
    \mathcal{P}_r^\textbf{P} \vert
    \mathbf{p}'\:  \mathbf{q}' \rangle_L
    =
    \frac{1}{2 h}
    \sum_g
    \chi_r ( g )
    (
    \delta_{R(g)\mathbf{p}, \mathbf{p}'} 
    \delta_{R(g)\mathbf{q}, \mathbf{q}'} + \delta_{R(g)(\mathbf{P}-\mathbf{p}), \mathbf{p}'} 
    \delta_{R(g)(\mathbf{P}-\mathbf{q}), \mathbf{q}'}  ).
\end{align}
where $\chi_r$ is the character for the irrep $r$ of group element $g$, and $R(g)$ are the three-dimensional rotation matrices for group element $g$, and $h$ is the order of the group (number of group elements). Notably, the characters for the trivial irrep, $A_1^+$ and $A_1$ are 1 for all elements. This projector operator is constructed to also satisfy the required symmetry for the identical particles in a boosted system. In constructing this operator, we ensure that it is hermitian, commutes with the transfer operator, and is idempotent. In proving these relations, we used the orthogonality theorems for characters, defined as
\begin{equation}
    \sum_{g} \chi_{i}(g)^* \chi_{j}(g) = h \delta_{ij}.
\end{equation}
Additionally, since the characters of the octahedral group and its associated little groups are strictly real, it is easy to show that we can then rewrite the evaluation of the transfer operator projected to irrep $r$,
\begin{align}
\label{eq:TandP}
     \langle \mathbf{p} \mathbf{q} \vert   \mathcal{P}_r\, \mathcal{T}\,  \mathcal{P}_r \vert \mathbf{p}' \mathbf{q}' \rangle_L =
     \langle \mathbf{p} \mathbf{q} \vert\mathcal{T}  \mathcal{P}_r \vert \mathbf{p}' \mathbf{q}' \rangle_L \,.
\end{align}
Evaluating the equality in Eq.~\eqref{eq:TandP} by inserting a complete set of momentum states, we can then express the projected transfer operator in this basis to ensure we compute energies only on irreps that couple to the scattering channel of interest.

\section{Charge radius derivation}
\label{app:radius}
The properties of a two-particle bound state are dominated by the long-distance asymptotic behavior of two non-interacting particles whenever its binding energy $B$ is small.
This universality was exploited in Ref.~\cite{10.1098/rspa.1935.0010} to explore the properties of the deuteron only using its binding energy as input, and neglecting the specifics of the short-distance interaction between the nucleons.
In this appendix, we derive the leading behavior of the charge radius of a two-particle bound state as a function of the binding momentum $\kappa = \sqrt{2\mu B}$.
First, we follow Ref.~\cite{10.1098/rspa.1935.0010}, deriving the charge radius from the solution of the non-relativistic Schrödinger equation.
In the second subsection, we derive this same behavior from the relativistic formalism used to extract the bound state form factor, i.e.\ taking the zero binding energy limit of Eq.~\eqref{eq:ffb}.

\subsection{Non-relativistic quantum mechanics}

The asymptotic radial behavior of a two-particle $s$-wave bound state wavefunction, assuming only an interaction of finite range $r_0$, is
\begin{equation}
    \psi(r\gg r_{0}) \sim \frac{e^{-\kappa r}}{r}\,,
\end{equation}
where $r$ is the relative distance between the two particles, $\kappa$ is the binding momentum $\kappa = \sqrt{2\mu B}$, $\mu$ is the reduced mass of the system and $B$ is the binding energy.
In the limit where $\kappa\to 0$, the asymptotic behavior will dominate the properties of the system.
In this limit, we can approximate the wavefunction everywhere using its asymptotic behavior, as originally derived in Ref.~\cite{10.1098/rspa.1935.0010}, with the normalized wave function equal to
\begin{equation}
    \psi(\mathbf{r}) = \sqrt{\frac{\kappa}{2\pi}} \frac{e^{-\kappa r}}{r}\,.\label{eq:asymtwobodywf}
\end{equation}
where $r=\abs{\mathbf{r}}$.

The charge radius of a two-body system, with one particle being charged and the other being neutral, is given by the expectation value of the location squared of the charged particle, which, for equal masses, corresponds to half of the relative distance,
\begin{equation}
    \ev{r_C^2} = \left\langle \left(\frac{r}{2}\right)^2\right\rangle
    = \frac{1}{4}\int d^3\mathbf{r}\, r^2 \abs{ \psi(\mathbf{r})}^2\,.\label{eq:chargeradfromwf}
\end{equation}
When plugging the wavefunction of Eq.~\eqref{eq:asymtwobodywf} into Eq.~\eqref{eq:chargeradfromwf} we obtain
\begin{equation}
    \ev{r_C^2} = \frac{1}{8\kappa^2}\,.
\end{equation}
This result diverges in the limit of zero binding energy, and as such will dominate the charge radius of states close to this limit, providing a universal behavior independent of the details of the finite range interaction.

\subsection{Relativistic formalism}
The form factor of the bound state system can be extracted with Eq.~\eqref{eq:ffb}, which we repeat here for the case of only one charged particle within the two-particle state
\begin{equation}
    f_B(Q^2) = \frac{g^2}{E_f+E_i}(\Ac^0(P_f,P_i)+f_1(Q^2)((E_f+E_i)\Gc(s_B,Q^2,s_B)-2\Gc^0(P_f,P_i)))\,.
\end{equation}

The bound state coupling $g^2$ given by Eq.~\eqref{eq:g2} can also be written as
\begin{equation}
    \frac{1}{g^{2}} = -\frac{\partial}{\partial s}(\rho\cot\delta -i\rho)_{s=s_B}\,.
\end{equation}
The first term is an analytic function of $s$ near threshold, while the second term, $\frac{\partial {\rho}}{\partial s}$, is proportional to $\kappa^{-1}$, which dominates this expression in the limit of zero binding energy, and leads to a vanishing coupling at $\kappa=0$.
In other words, the leading behavior of the coupling near threshold is proportional to $\kappa$,
\begin{equation}
\label{eq:g2kinZB}
    g^2 = \left(\frac{\partial }{\partial s}i\rho\right)^{-1}_{s=s_B} + \mathcal{O}(\kappa^2)\,.
\end{equation}

Similar to the behavior of $g^2$ driven by the non-analytic part of the scattering near threshold, the behavior of $f_B$ is driven by the non-analytic contribution of the triangle contribution, dominating over the analytic contribution from $\Actwo^0$.
In particular, the anomalous singularity of the triangle diagram located at the timelike momentum
\begin{equation}
    Q_A^2 = -4 s_B  \left( 1 - \frac{s_B}{4 M^2}\right)\,.
\end{equation}
moves from $Q^2<0$ towards $Q^2=0$ as $\kappa\to 0$.
This divergent singularity combined with the vanishing coupling gives rise to a finite and universal form factor as $\kappa\to0$, which is independent of the dynamics of the system described by the phaseshift $\delta$ and the smooth function $\Actwo^0$,
\begin{equation}
f_{B,\text{tri.}}(Q^2)
    = \frac{1}{E_f+E_i}\left(\frac{\partial }{\partial s}i\rho\right)^{-1}_{s=s_B}
    f_1(Q^2)((E_f+E_i)\Gc(s_B,Q^2,s_B)-2\Gc^0(P_f,P_i))
    + \mathcal{O}(\kappa)\,.
\end{equation}
In particular, it can be shown that in the limit of $\kappa\to0$ this form factor has the correct normalization at $Q^2=0$
\begin{equation}
    \lim_{\kappa\to 0}f_{B,\text{tri.}}(0) = 1\,.
\end{equation}

We can calculate the charge radius associated with this form factor to obtain
\begin{equation}
    \ev{r_C^2}_{\text{tri.}}=-6\frac{\mathrm{d}}{\mathrm{d} Q^2} f_{B,\text{tri.}}(Q^2)\big \vert_{Q^2=0} = \frac{1}{8\kappa^2} +\mathcal{O}(1)\,,
\end{equation}
showing the same universal dominating result as in the quantum mechanical case of the previous section.
\end{document}